\documentclass[iop,apj,twocolappendix,appendixfloats,numberedappendix]{emulateapj}
\usepackage{amsmath,amssymb,amstext}

\usepackage[breaklinks,colorlinks,citecolor=blue,linkcolor=magenta]{hyperref} 

\usepackage[all]{hypcap} %Links go to figures; breaks on deluxetables (use \capstartfalse \capstarttrue to fix it)

\usepackage{graphicx}
\usepackage{aas_macros}
\usepackage{natbib}
\usepackage{mathtools}
\usepackage{cases}
\usepackage{soul}
\usepackage[usenames,dvipsnames]{xcolor}
\usepackage[colorinlistoftodos]{todonotes}

\newcommand{\rp}{\emph{r}-process}
\newcommand{\g}{\ensuremath{\gamma}}
\newcommand{\al}{\ensuremath{\alpha}}

\newcommand{\bt}{\ensuremath{\beta}}
\newcommand{\Msun}{\ensuremath{M_{\odot}}}
\newcommand{\mej}{\ensuremath{M_{\rm ej}}}
\newcommand{\vej}{\ensuremath{v_{\rm ej}}}
\newcommand{\dd}[1]{\ensuremath \mathrm{d} #1}
\newcommand{\ftot}{\ensuremath{f_{\rm tot}(t)}}
\shorttitle{Kilonova Thermalization}
\shortauthors{Barnes et al.}

\begin{document}

\title{Radioactivity  and thermalization in the ejecta of compact object
  mergers and their impact on kilonova light curves} 
\author{Jennifer Barnes$^{1}$, Daniel Kasen$^{1,2}$, Meng-Ru Wu$^4$,
  Gabriel Mart{\'i}nez-Pinedo$^{3,4}$} 

\altaffiltext{1}{Departments of Physics and Astronomy, 366 LeConte
  Hall, University of California, Berkeley, CA, 94720} 
\altaffiltext{2}{Nuclear Science Division, Lawrence Berkeley National
  Laboratory, Berkeley, CA 94720} 
\altaffiltext{3}{GSI Helmholtzzentrum f{\"u}r Schwerionenforschung,
  Planckstra{\ss}e 1, 64291 Darmstadt, Germany} 
\altaffiltext{4}{Institut f{\"u}r Kernphysik (Theoriezentrum),
  Technische Universit{\"a}t Darmstadt, Schlossgartenstra{\ss}e 2,
  64289 Darmstadt, Germany} 
\email{jlbarnes@berkeley.edu}

\begin{abstract}
 One of the most
  promising electromagnetic signatures of compact object mergers are
  kilonovae: approximately isotropic radioactively-powered transients
  that peak days to weeks post-merger. Key uncertainties in modeling
  kilonova light curves include the emission profiles of the radioactive decay products---non-thermal beta-particles, alpha-particles, fission
  fragments, and gamma-rays--- and the efficiency with which they deposit their energy in the ejecta.  The amount of radioactive energy and the efficiency of its thermalization sets the luminosity budget and is
  therefore necessary for predicting kilonova light curves. We outline the uncertainties in \rp\ decay, describe
  the physical processes by which the kinetic energy of the decay products is absorbed in the
  ejecta, and calculate time-dependent thermalization
  efficiencies for each particle type. We determine the net thermalization efficiency and
  explore its dependence on \rp\ nucleosynthetic yields---in
  particular, the production of translead nuclei that undergo
  \al-decay---and on the ejecta's mass, velocity, composition, and
  magnetic field configuration. We incorporate our results into new
  time-dependent, multi-wavelength radiation transport simulations,
  and calculate updated predictions of kilonova light curves.
Thermalization has a substantial effect on kilonova
photometry, reducing the luminosity by a factor of roughly 2 at peak, and
by an order
  of magnitude or more at later times (15 days or more after explosion). We present simple analytic fits to time-dependent
  thermalization efficiencies, which can easily be used to improve light
  curve models.  We briefly revisit the putative kilonova that accompanied
  gamma ray burst 130603B, and offer new estimates of the mass ejected
  in that event. We find that later-time kilonova light curves
can be significantly impacted by \al-decay from translead isotopes; data at these times may therefore
be diagnostic of ejecta abundances.
\end{abstract}

\maketitle

\section{Introduction}

In addition to producing kilohertz gravitational waves (GW) detectable
by ground-based interferometers \citep{Abadie_2010_LIGO}, compact
object (CO) mergers involving a neutron star (NS) are likely to emit a
variety of electromagnetic (EM) signals. Immediately post-merger, the
accretion of disrupted NS material onto the central black hole (BH) or
hypermassive NS may drive a short gamma ray burst
\citep{Paczynski_1986_GRB,Eichler_1989_sGRBs,Narayan_1992_sGRBs}. Mergers
may also produce optical/infrared transients
\citep{Li_Paz_1998,Metzger_2010,Roberts_2011,Barnes_2013} powered by
the radioactive decay of heavy elements synthesized via rapid neutron
capture~\citep[the \rp;][]{arnould.goriely.takahashi:2007}.  The \rp~
is expected to operate in material ejected from the system
dynamically~\citep{Lattimer_1974,Lattimer_1976,Freiburghaus_1999,Korobkin_NSM_rp,Rosswog_1999,Goriely_2011},
or unbound from a remnant accretion
disk~\citep{Fernandez_Metz_2013_diskOutlfows,Perego_2014_vWinds,Just_2015_torusNucleo}.
On much longer timescales, the interaction of the ejecta with the
interstellar medium will generate a radio
signal~\citep{Nakar_Piran_2011_radio}.

Observing an EM counterpart will enhance the science returns of a GW
detection \citep{Metzger_Berger_2012} by identifying the host galaxy
and the position of the merger within the host
\citep{Nissanke_2013,Kasliwal_Niss_2014,Holz_Hughes_2005_GWEM,
  Dalal_etal_2006}, constraining the neutron star equation of state
\citep{Bauswein_2013,Hotokezaka_2013,Bauswein_2015}, and confirming
low signal-to-noise GW events \citep{Kochanek_1993,
  Harry_Fairhurst_followup}.  Among possible counterparts, the radioactive
transients---known as ``kilonovae''---are especially
promising. Kilonova emission is roughly isotropic
\citep{Roberts_2011,Bauswein_2013}, and peaks on timescales of
days--weeks post-merger
\citep{Barnes_2013,Tanaka_Hotok_rpOps,Grossman_2014_kNe}, making it
ideal for EM follow-up of a GW trigger.  Because kilonovae derive
their energy from radioactive decay, they probe nucleosynthesis in the
merger in a way other counterparts cannot, and may therefore constrain
the astrophysical origin of \rp\ element production.

Accurate models of kilonova photometry are crucial for dual detection
efforts. Unfortunately, the exotic composition of the heavy element
ejecta, and uncertainties in \rp\ nucleosynthesis and decay, pose
challenges to radiation transport simulations required to build these
models. Recent work \citep{Kasen_2013_AS} clarified the opacity of
\rp\ material, reducing a major uncertainty in kilonova radiation
transport simulations, but other key inputs remain relatively
unconstrained.

Any rigorous
  kilonova model must address the following aspects of radioactivity: \emph{i})
  the total amount of radioactive energy released; \emph{ii}) the
  decay channels that dominate the energy production during different
  phases of kilonova evolution; and \emph{iii}) the efficiency with which
suprathermal radioactive decay products--- \bt-particles,
\al-particles, \g-rays, and fission fragments---transfer their energy
to the thermal background.  Once thermalized, the energy is
re-radiated as thermal emission, powering the kilonova light
curve. 

Although thermalization determines the kilonova's luminosity,
and is thus essential for light curve modeling, no detailed
calculation of thermalization efficiencies has been attempted.
\citet{Metzger_2010} presented analytic estimates of thermalization,
but focused on timescales shorter than those now believed to
characterize kilonova light curves.  \citet{Hotokezaka_2015_heating}
studied \g-ray deposition in kilonovae, but did not investigate the
thermalization of charged particles, which carry a large fraction
of the radioactive energy.

Modeling the thermalization of \rp\ decay energy in the kilonova ejecta is
challenging.  Thermalization rates are sensitive to the ejecta's mass,
velocity, and composition, as well as its magnetic field structure,
which has not been definitively determined by magnetohydrodynamic
simulations.  The broad range of elements synthesized by the \rp, and
the often unknown properties of the heaviest of those elements,
complicates the situation, as does the complexity of the net emission
spectrum, which is a sum over several decay chains, each evolving on
its own timescale.

This paper addresses the issues outlined above, with special emphasis
on the key physical processes influencing the
thermalization of \rp\ decay products in kilonovae.  In \S \ref{sec:ejMod}, we describe our ejecta model and
its uncertainties. Section \ref{sec:thermRates} defines energy loss
rates for \bt-particles, $\alpha$-particles, $\gamma$-rays, and
fission fragments, and explores their sensitivity to ejecta
parameters. Analytic estimates and analytic expressions for
thermalization efficiencies are developed in \S
\ref{sec:analytics}. In \S \ref{sec:results}, we present detailed numerical calculations of
time-dependent thermalization efficiencies $f(t)$ for individual
species and 
for the system as a whole, and discuss the sensitivity of
$f(t)$ to properties of the ejecta. Finally, \S \ref{sec:LCs}
evaluates the effect of thermalization on kilonova light curves, and
uses improved light curve models to estimate the mass ejected by the claimed
kilonova associated with GRB 130603B.

\section{Properties of the kilonova ejecta}
\label{sec:ejMod}
\subsection{Ejecta model}

Predictions of kilonova outflows vary, due to natural diversity in the
merging systems (e.g. different mass ratios, BHNS v. NS$^2$) and
uncertainties in the NS EOS. Recent hydrodynamic simulations
\citep{Bauswein_2013,Hotokezaka_2013, Kyutoku_2015_massEj,
  Sekiguchi_2105_massEj} suggest that mergers dynamically eject
between $\sim 10^{-4}$ and a few $\times 10^{-2}$ \Msun\ of material, with bulk velocities of a few tenths the speed of
light. Additional material ($\sim 10^{-3}-10^{-2} \Msun$) can exit the
system at slightly lower velocities ($0.05c - 0.1c$) as a wind from an accretion torus
\citep{Fernandez_Metz_2013_diskOutlfows,Perego_2014_vWinds,
  Fernandez_etal_2015_BHspin,Just_2015_torusNucleo}.

We adopt as our fiducial model a system with \mej = $5 \times 10^{-3}$
\Msun\ and \vej = $0.2c$, where \vej\ is defined in terms of the
explosion kinetic energy, $E_{\rm k} = \mej\vej^2/2$. Since denser
ejecta configurations thermalize more efficiently than diffuse
systems, we vary these parameters over the ranges
$\mej/\Msun \in [10^{-3}, 5\times 10^{-2}]$ and
$\vej/c \in [0.1, 0.3]$.

We assume the ejecta is spherical and expanding homologously, and that
the density profile follows a broken power-law, declining with
velocity coordinate $v = r/t$ as $v^{-\delta}$ in the inner regions of
the ejecta, and $v^{-n}, \: n > \delta$, in the outer regions. We set
$\delta=1$ and $n=10$. \citet{Barnes_2013} provides a complete
mathematical description of the density profile.

\subsection{Magnetic fields}
\label{subsec:Bej}
Kilonova ejecta contain a residual magnetic field, either inherited
directly from the parent neutron star(s), or seeded by amplified
fields produced by turbulence during the merger or in the resultant accretion disk
\citep{Kiuchi_2014_BAmp,Kiuchi_2015_BAmp}.
Though weakened by expansion, the fields remain strong enough to
influence charged particle motion. In a sufficiently strong field,
charged particles have Larmor radii smaller than the coherence length
of the magnetic field, and their motion is confined to ``flux tubes''
that trace the field lines.

If magnetic flux is frozen into the homologously-expanding ejecta, the field strength is related to the ejecta radius $R_{\rm ej} = \vej t$ by
\begin{align}
B(t) \approx \frac{B_{0} R_{0}^2}{R_{\rm ej}^2} 
\approx 3.7 \times 10^{-6} B_{12} R_6^2 v_2^{-2} t_{\rm d}^{-2} \mbox{ G},
\end{align}
where $v_2 = v/0.2c$, $t_{\rm d}$ is the elapsed time in days, and
$B_0$ and $R_0$ are the magnetic field and radius at the time of mass
ejection. The quantities $B_{12} = B_0/10^{12}~$G and
$R_6 = R_0/10^6~$cm have been scaled to typical values;
$R_0 \approx 10^6 - 10^7$ cm is characteristic of the size of NSs or
the post-merger accretion disk, and $B_0$ may range from
$10^9 - 10^{15}$~G, depending on the initial NS fields and the
efficiency of magnetic field amplification.

A relativistic particle of mass $m$, charge $q$, kinetic energy $E$,
and velocity $v$ in a magnetic field $B$ has a maximum Larmor radius
(when $\mathbf{v} \perp \mathbf{B}$) of
\begin{equation*}
r_{\rm L,max} = \frac{(E + mc^2) v}{qBc},
\end{equation*}
Assuming typical emission energies of $E_{\bt,0} = 0.5$ MeV for
\bt-particles, $E_{\al,0} = 10$~MeV for \al-particles, and
$E_{\rm ff,0} = 150$ MeV for fission fragments, and assuming fission
fragments are singly ionized and have masses of $\sim 130~m_{\rm u}$, with $m_{\rm u}$ the
atomic mass unit, the Larmor radii are
\begin{align}
\frac{r_{\rm L,max}(t)}{R_{\rm ej}(t)} =
\begin{cases}
1.5 \times 10^{-6}~ v_2 t_{\rm d} B_{12}^{-1}R_6^{-2} &\mbox{ \bt-particles} \\
2.4 \times 10^{-4}~ v_2 t_{\rm d} B_{12}^{-1}R_6^{-2} &\mbox{ \al-particles} \\
1.0 \times 10^{-2}~ v_2 t_{\rm d} B_{12}^{-1}R_6^{-2} &\mbox{ fiss. fragments}.
\end{cases}
\end{align}

We will adopt the flux tube approximation for all particles. This is
clearly appropriate for \al- and \bt-particles, which have
$r_{\rm L}/R_{\rm ej} \ll 1$ for the duration of the kilonova. Fission
fragments may, at later times, have $r_{\rm L}$ large enough that they
jump from field line to field line.  We discuss this possibility in
\ref{subsec:fragTrans}, but do not employ models of fission fragment transport beyond the flux tube approximation in this work.

The magnetic field structure determines charged particle
trajectories and so affects thermalization.  Radial fields can escort
fast charged particles straight out of the ejecta, reducing
thermalization. In contrast, toroidal or tangled fields trap charged
particles, and so enhance thermalization. We consider three types of
configurations here: radial ($\mathbf{B} \propto \hat{\mathbf{r}}$),
which may be produced by the outward motion of the ejecta ``combing''
out the field lines; toroidal
($\mathbf{B} \propto \hat{\boldsymbol{\phi}}$), perhaps created by the
spiral motion of neutron stars shedding mass through tidal stripping;
and random, which may be generated by turbulent motions in the
material during mass ejection. To model the latter case, we assume
field lines re-orient on a length scale $\lambda R_{\rm ej}$, where
the dimensionless parameter $\lambda < 1$.
  
\subsection{Composition}
\label{subsec:comp}

The ejecta composition impacts thermalization in two ways.  First, it
determines the partition of radioactive energy among different decay
channels, and the energy spectra of the decay products.  Second,
it sets the properties of the background material (e.g., isotope mass,
atomic number, and ionization energy) which influence the energy loss
rates of the decay products.

To determine the ejecta composition, we calculate \rp\ nucleosynthesis
on a set of smoothed-particle hydrodynamics (SPH) trajectories
extracted from a relativistic simulation of an equal-mass (1.35
\Msun-1.35 \Msun) NS$^2$ merger \citep{Goriely_2011}.  At the start of
the nucleosynthesis calculation, all trajectories had temperatures of
6 GK, densities set by their hydrodynamical histories, compositions
determined by nuclear statistical equilibrium, and initial electron
fractions $Y_{\rm e,0}$ between $1.5 \times 10^{-2}$ and
$5.5 \times 10^{-2}$. These quantities were then evolved according to
the reaction network, which tracks charged particle reactions, neutron
capture, photo-dissociation, \bt- and \al-decay, and fission, for more
than 7300 nuclei. \citep[For a detailed description of the network, see][]{MendozaTemis_etal_rProcess}.

The hydrodynamical model predicts two classes of trajectories that
produce two distinct compositions: ``slow'' trajectories, where all
free neutrons are depleted by neutron-capture, and ``fast''
trajectories, where early rapid expansion precludes the capture of all
neutrons by seed nuclei~(see also: \citealt{Just_2015_torusNucleo,
  Goriely_2014_rP_fast,
  Metzger_2015_nPrecurs,MendozaTemis_etal_rProcess}).  The slow
trajectories comprise $\sim 90$\% of the ejecta, and robustly produce
\rp\ elements up to the third peak. The fast ejecta \textit{r}-pattern
is different from the slow component due to the longer neutron-capture
timescale. 
In constructing our ejecta model, we assume that at times relevant for thermalization, material from slow trajectories will be located in the ejecta's interior regions (henceforth ``inner ejecta''), while material from the fast trajectories occupies the outer regions (``outer ejecta'').
We sum the trajectories in each
class to construct mass-integrated inner and outer compositions.  We
also select a representative case from the set of inner trajectories,
which typifies conditions in the merger ejecta. We use this
representative trajectory to study the details of the radioactivity.

\begin{figure}
\includegraphics[width=3.5 in]{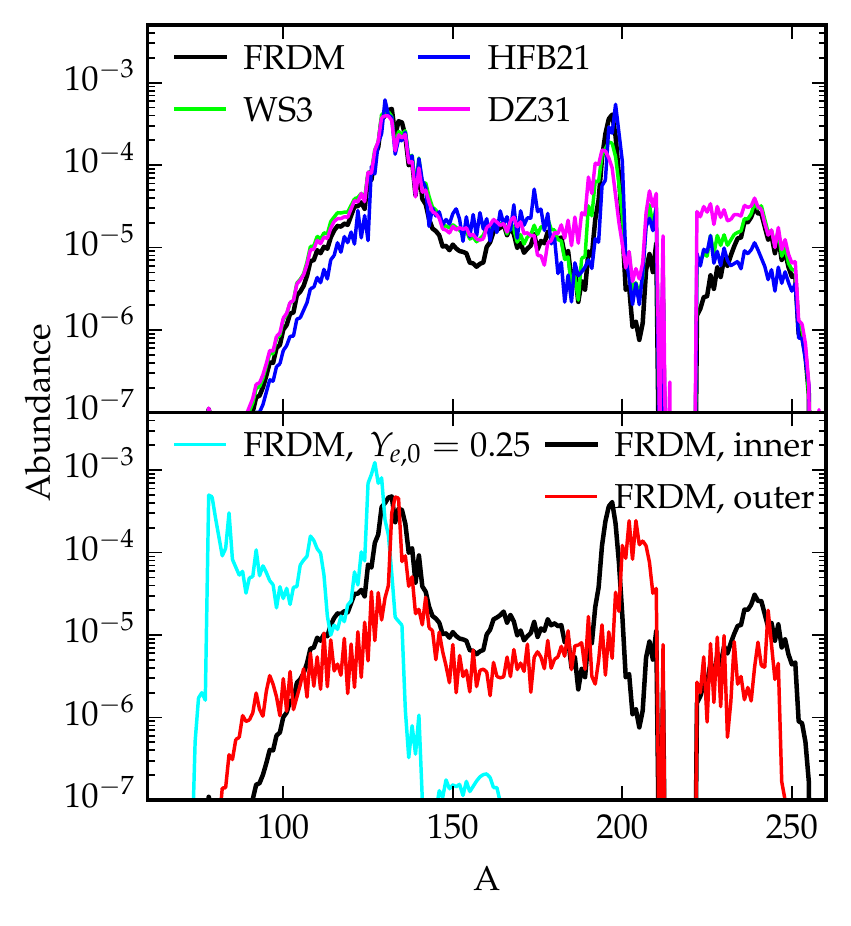}
\caption{ Abundance yields from our nuclear network calculations at $t
  = 1$ day. \textit{Top panel:} Mass-integrated abundances from the
  ``inner'' ejecta ($\sim 90$\% of the ejected mass) for four nuclear
  mass models. The \rp\ proceeds past the third peak, and strong
  fission cycling reduces differences among nuclear mass
  models. \textit{Bottom panel:} An illustration of major factors
  affecting the final abundances. The red curve shows mass-integrated
  abundances for the ``outer'' ejecta, using the FRDM mass
  model. Rapid expansion hinders free neutron capture, decreasing
  heavy element production, and creating a substantial amount of H as
  uncaptured neutrons decay to protons. The cyan curve shows the
  abundance yield of a representative inner ejecta trajectory, whose
  initial electron fraction has been artificially increased to $Y_{\rm e,0} =
  0.25$, leading to limited production of nuclei with $A > 130$.} 
\label{fig:comp}
\end{figure}

The top panel of Figure \ref{fig:comp} shows the inner ejecta
composition at $t=1$ day, based on neutron-capture and
photodissociation rates computed using the statistical model for
four different nuclear mass models: the Finite Range Droplet Model
(FRDM; \citealt{Moller_1995_FRDM}), the Hartree-Fock-Bogoliubov model
HFB21 \citep{Goriely_HFB21}, the Weizs{\"a}cker-Skyrme model (WS3;
\citealt{Liu_etal_2011_WS3}), and the Duflo-Zucker model with 31
parameters (DZ31; \citealt{Duflo_Zuker_1995_DZ31}). We find that
although the abundance pattern is similar for different mass models,
particularly around $A \sim 130$ due to fission cycling, the position
of the peak at $A\sim 195$ and the abundances for $A \gtrsim 195$ depend on the mass
model. The differences in the translead abundances impact the late time kilonova light curves, 
as will be discussed in  \S \ref{subsec:lateLC}.

The abundances are also influenced by the electron fraction
$Y_{\rm e,0}$ at the onset of the \rp; neutron scarcity (high
$Y_{\rm e,0}$) suppresses the assembly of the heaviest \rp\ elements.
Our SPH trajectories are all initially very neutron rich; however,
weak interactions in the aftermath of a merger could raise
$Y_{\rm e, 0}$
substantially~\citep{Wanajo.Sekiguchi.ea:2014,Sekiguchi.Kiuchi.ea:2015,Goriely.Bauswein.ea:2015}.
To explore this effect, we
artificially increased the initial $Y_{\rm e,0}$ of our
representative trajectory from its value of $0.04$, and reran the nuclear reaction network.  As
expected, higher initial electron fractions produce fewer heavy
elements (bottom panel of Figure \ref{fig:comp}) and for
$Y_{\rm e,0} \gtrsim 0.2$ the \rp\ fails to reach the third peak,
instead producing material with $A \sim 70 - 110$.

The ejecta composition evolves with time as neutron-rich isotopes
gradually decay to stability.  However, on timescales relevant for
kilonova light curves ($t \sim 0.1 - 10$ days), this evolution is
fairly slow, and driven primarily by \al- and \bt-decays, which do not
dramatically change the abundance-averaged properties of the composition.  For the
purpose of calculating energy loss rates, we therefore assume that the
abundance-averaged properties are constant in time, 
but vary in space, with the inner 90\% (outer 10\%) of the mass
described by the average abundance pattern of the inner (outer)
trajectories, calculated at $t=1$ day using the FRDM mass model.  The
outer ejecta differs from the inner ejecta primarily in its high
abundance of hydrogen, produced by the decay of remnant free neutrons to protons.

\subsection{Radioactivity}
\label{subsec:Ej_Radio}

The energy generation rate from r-process decay has been shown to
approximately follow
$\dot{\epsilon} = \epsilon_0 t_{\rm d}^{-\alpha}$, with
$\epsilon_0 \approx 10^{11} \text{ ergs s}^{-1} \text{g}^{-1}$
and $\alpha = 1.1 - 1.4$
\citep{Metzger_2010,Roberts_2011,Goriely_2011,Korobkin_NSM_rp}.
However, the fraction of the energy supplied by each decay channel,
and the emission spectra for each decay product, are less clear.

Though \rp\ radioactivity is most commonly associated with
  \bt-decay, any translead nuclei synthesized will decay by \al-emission, and heavier ($A \gtrsim 250$) nuclei may undergo fission,
  trends which have implications for thermalization.  Radioactive
  emission in kilonovae will at all times be dominated by isotopes
  with half-lives $\tau_{1/2}$ of order $t_{\rm exp}$, the time since explosion. For any
  particular decay channel, $\tau_{1/2}$ is strongly correlated with
  the energy $Q$ emitted when a nucleus decays. However, this is not
  true across decay channels; for a given $\tau_{1/2}$, \bt-decay has
  lower $Q$ than \al-decay, which has lower $Q$ than fission.  As a
  result, \al-decay can generate a substantial fraction of the \rp\
  radioactive energy, even though \rp\ yields are dominated by
  nuclei that \bt-decay.  Fission could, in principle, also be an
  important source of energy, but we find almost all fissioning nuclei
  have $\tau_{1/2}$ less than a day, suggesting that fission supplies
  a negligible amount of energy after very early times.  Since energy
  from \al-decay thermalizes with a different efficiency than
  \bt-decay energy, thermalization depends on the relative importance of these decay channels, and thus on the yields of
  translead nuclei.

\begin{figure}
\includegraphics[width=3.5 in]{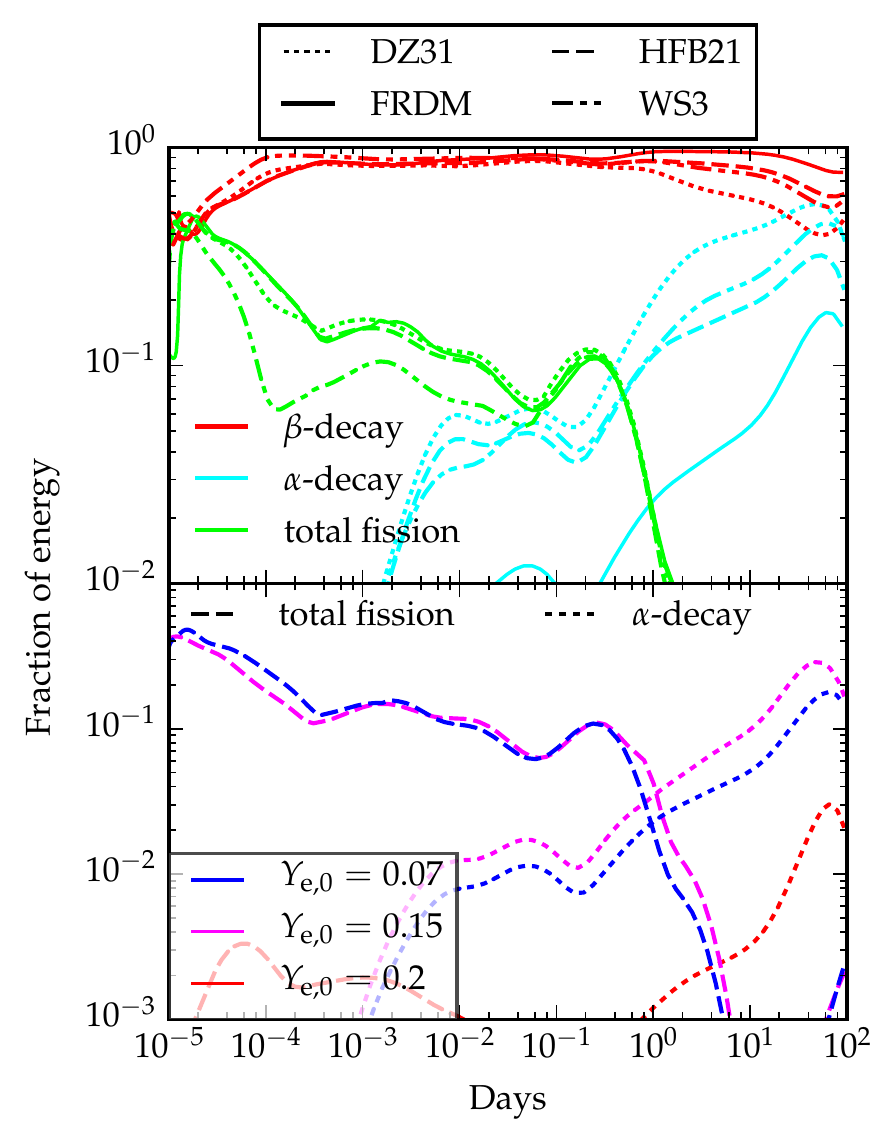}
\caption{ \textit{Top panel:} The fraction of the total radioactive energy
  produced by \bt-decays, \al-decays, and fission in our
  representative trajectory, for four nuclear mass
  models. While \bt-decay dominates, fission (\al-decay) can be
  important at early (late) times. Agreement between the four mass
  models studied is within an order of magnitude.  
\textit{Bottom panel:}  Energy released in \al-decays and fission, for
the FRDM mass model, for a range of $Y_{\rm e,0}$. Lower electron
fractions favor the assembly of the heavy elements that later decay
by fission and \al-emission. As $Y_{\rm e,0}$ increases, these
processes become less important, and are negligible for $Y_{\rm e,0} >
0.2$.} 
\label{fig:en_gen}
\end{figure}

The top panel of Figure \ref{fig:en_gen} shows the fraction of radioactive
energy produced by \al-decay, \bt-decay, and fission, for the
representative trajectory introduced in \ref{subsec:comp}, calculated
for four nuclear mass models. Beta-decay is the primary source of
energy for all mass models out to late times. Fission, including
\bt-delayed, neutron-induced, and spontaneous fission, contributes
$\sim 10$\% of the energy at times $\lesssim 1$ day, and \al-decay
becomes significant within a few hours. The fractions for the
different nuclear mass models generally agree with each other, but the
estimates of energy generated by \al-decay differ by a factor of
almost ten, with DZ31 predicting the largest
    contribution from \al-decay and FRDM predicting the least. Since
  \al-decays release more energy, per decay, than \bt-decay, the
  enhanced role of \al-decay predicted by the DZ31 model also results
  in an increase in the total energy generated by the decay of \rp\ isotopes. The
  increase is modest early on (a factor of $\lesssim 1.2$ for
  $t \leq 1$ day) but becomes more important at late times (a factor
  of $\gtrsim 2$ by $t =$ 1 month.)

The bottom panel of of Figure~\ref{fig:en_gen} explores the effect of
electron fraction on the decay channels.  Fission and \al-decay are
significant sources of energy for $t\lesssim 1$~day and
  $t\gtrsim 1$~day, respectively, for ejecta with
$Y_{\rm e,0} \lesssim 0.2$, but become negligible at higher
$Y_{\rm e,0}$ because the reduced number of free neutrons chokes the
production of the heaviest nuclei.

\subsection{Emission Spectra of Decay Products}
\label{subsec:spec}

Modeling the energy spectra of \rp~radioactive decay products is
complicated by the large number of decay chains and the uncertain
nuclear data. However, we can construct approximate spectra by
considering emission from a range of contributing decays. We calculate
emission spectra using the time-dependent composition of our
representative inner SPH trajectory.  The decay energies for \al- and
\bt-decay were determined from experimental mass excesses (AME 2012;
\citealt{AME2012_a, AME2012_b}) when available, and theoretical (FRDM)
mass excesses otherwise.  Decay data, including \bt\ endpoint energies,
\g-spectra, and half-lives for \bt- and \al-decay, were retrieved from
the Nuclear Science References database \citep{NucData_References},
accessed via the website of the International Atomic Energy Agency.

\subsubsection{Beta decay}

Energy from \bt-decay takes the form of energetic \bt-particles,
\g-rays, and neutrinos. (Beta-delayed fission is treated as
  part of fission, and we neglect \bt-delayed neutron- and
  \al-emission, as they are expected to be negligible for nuclei with
  lifetimes longer than a day.) Following \bt-emission, nuclear
de-excitation can also emit low-energy atomic electrons, delayed
neutrons, and $\sim$ keV X-rays, but we found that these secondary
processes were negligible.

We constructed \bt- and \g-spectra using selected isotopes
that dominated the \bt-decay energy production.  The energy generation
rate of an isotope $i$ was estimated as
$\dot{\epsilon}_{\rm \bt,i} = Y_{\rm i} Q_{\rm \beta,i}/\tau_{\rm 1/2,i}$,
where $Y_{\rm i}$ is the number abundance of the isotope,
$Q_{\rm \beta,i}$ the decay energy, and $\tau_{\rm 1/2,i}$ the half
life.  We used experimental values for $Q_{\rm \beta}$ and
$\tau_{\rm 1/2}$ when available, and theoretical values otherwise.
We excluded isotopes lacking decay data and those with heating rates
less than 1\% of the maximum single-isotope heating rate. The excluded
\bt-decays account for only 5-7\% of the total \bt-decay energy at all
times. The \g-ray intensities were taken directly from nuclear
measurements, while $\beta$-spectra were constructed from endpoint
energies and intensities assuming all decays had an allowed spectral
shape and using the simplified Fermi formula fit proposed by
\citet{Schenter_Vogel_1983_betaSpec}.

\begin{figure}
\includegraphics[width = 3.5 in]{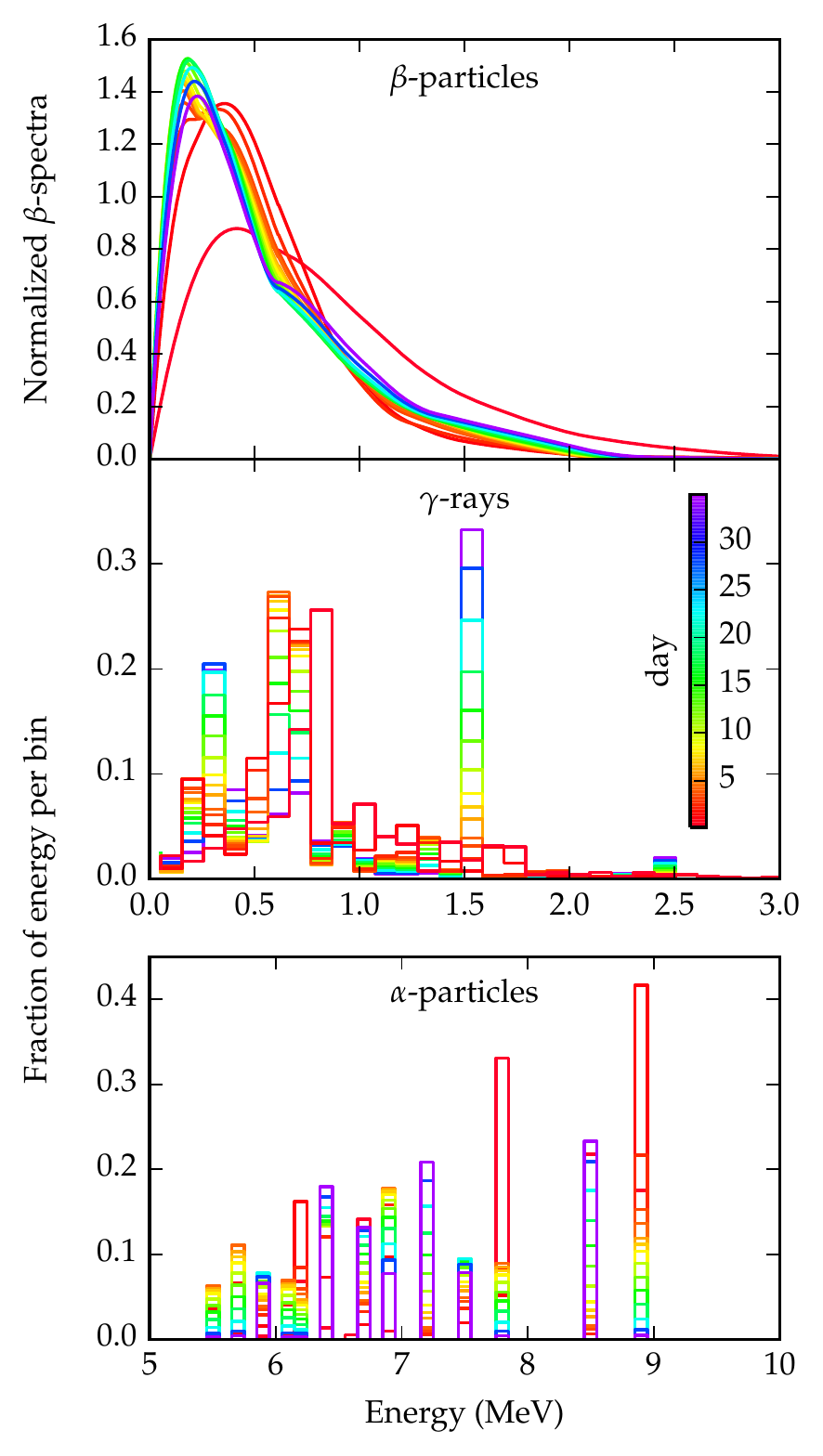}
\caption{The emission spectra for \bt-particles (top panel), \g-rays (middle panel), and \al-particles (bottom panel) as a function of time.}
\label{fig:netSpec}
\end{figure}

We find that roughly $20\%$ of the \bt-decay energy emerges as
\bt-particles, $45\%$ as \g-rays and 35\% as neutrinos.  The energy
lost to neutrinos, which escape the ejecta without depositing any
energy, sets an upper limit of $\sim 65 \%$ on the \bt-decay
thermalization efficiency.  The top two panels of
Figure~\ref{fig:netSpec} show the \bt- and \g- spectra for the composition of our neutron-rich representative SPH trajectory for $t = 1$ -- $30$ days. The
\g-ray spectra peak at several hundred keV and the \bt-spectra at
around 0.5 MeV.

The \bt-spectrum was found to be consistent across mass models, which is not surprising since \bt-decay energy is very sensitive to half-life, and \bt-emission at all times is dominated by nuclei with half-lives of order the time since explosion. However, we did find the spectrum depends mildly on electron fraction, with higher $Y_{\rm e,0}$ slightly enhancing the spectrum's high-energy tail.

\begin{figure}
\includegraphics[width = 3.5 in]{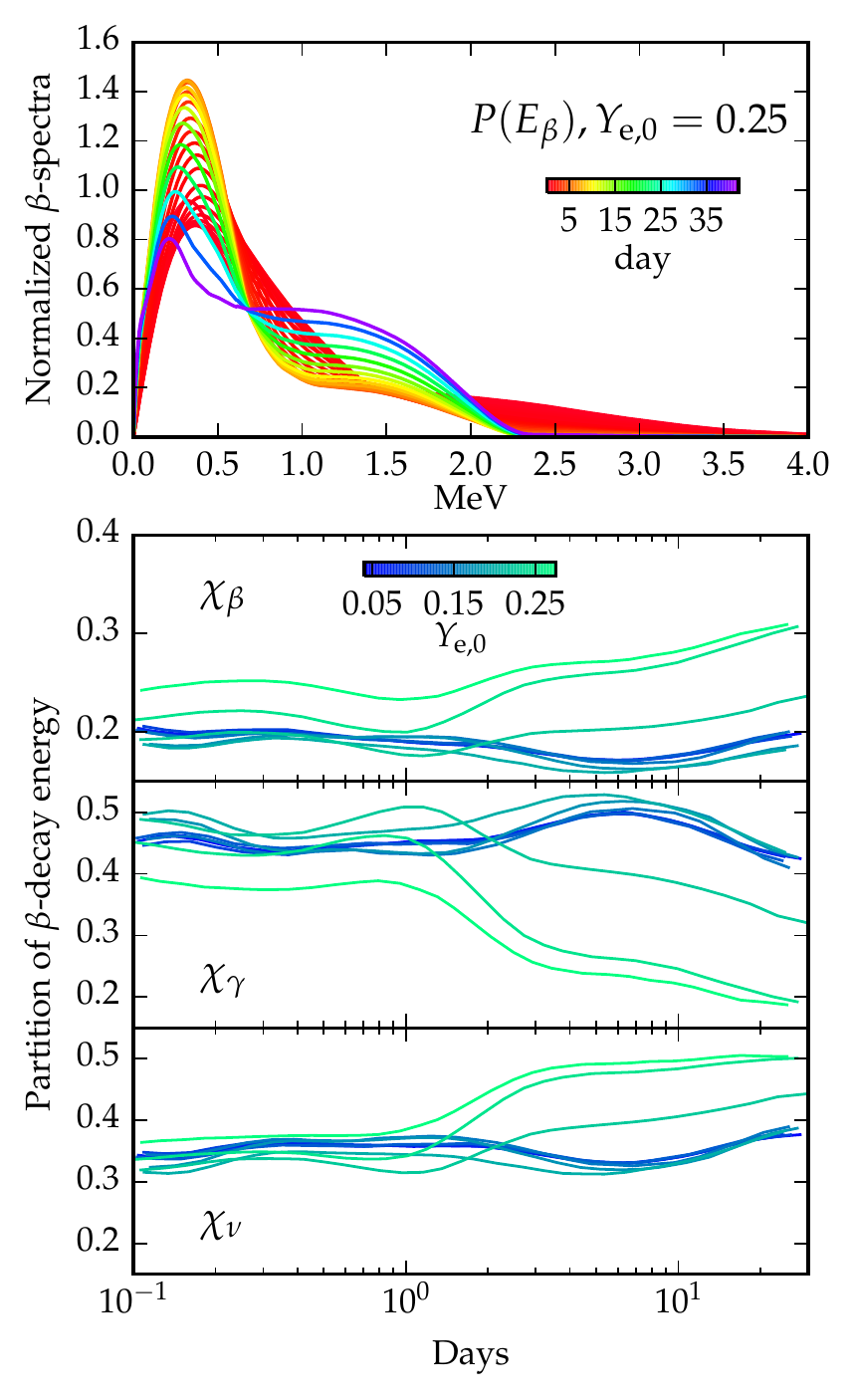}
\caption{The effect of electron fraction $Y_{\rm e,0}$ on \bt-decay
  for the FRDM nuclear mass model. (Other nuclear mass models studied
  showed similar trends). \textit{Top panel:} The \bt-spectrum for a
  composition with $Y_{\rm e,0}= 0.25$. The spectrum is shifted to
  higher energies, relative to the low-$Y_{\rm e,0}$ case
  (Figure~\ref{fig:netSpec}). \textit{Bottom panels:} The fraction of
  $Q_{\beta}$, $\chi$, imparted to \bt-particles, \g-rays, and neutrinos, for
  different values of $Y_{\rm e,0}$. As $Y_{\rm e,0}$ increases, a
  greater fraction of $Q_{\bt}$ goes to \bt's and neutrinos, while
  $\chi_{\g}$ shrinks. This effect is particularly pronounced at late
  times. }
\label{fig:YeDep}
\end{figure}

This is due to differences in how \bt-decay energy is
  divided among \bt-particles, \g-rays, and neutrinos. Compositions evolved
  from higher initial $Y_{\rm e}$ impart a greater \emph{fraction} of the
  total \bt-decay energy $Q_{\bt}$ to \bt-particles, at the expense of
  \g-rays (see the lower three panels of Figure \ref{fig:YeDep}).  As
  shown in Figure~\ref{fig:comp}, higher electron fractions yield
  compositions with lower $A$.  The \bt-decays for these lighter
  nuclei tend to be dominated by one or a few transitions to low-lying
  nuclear energy states; the energy carried away by the \bt-particle
  and the neutrino is close to $Q_{\bt}$, and the energy released in
  \g-rays during nuclear de-excitation is reduced. In contrast, for
  more massive nuclei, the excitation energy of the daughter nucleus
  after emission of the \bt-particle and neutrino is more likely to be
  a significant fraction of $Q_{\bt}$, and a greater portion of the
  energy takes the form of \g-rays.  Therefore, despite having similar
  $Q_\bt$, nuclei synthesized in high-$Y_{\rm e,0}$ conditions
  generate more energetic \bt-particles.  We found these effects to be
  independent of mass model.

\subsubsection{Alpha decay}

While the majority of species produced by the \rp\ stabilize through
\bt-decay, some heavier isotopes ($A \gtrsim 200$) undergo
\al-decay. Unlike \bt-particles, \al-particles are ejected from nuclei
at discrete energies that fall within the fairly narrow range
$E_{\al} \sim 5 - 9$ MeV. Due to the fact that alpha decay is a
tunneling process, \al-particles carry all of the decay energy in the
majority of decays, and the incidence of \g-emission is vanishingly
low.

We determined the most important sources of \al-emission using the
procedure detailed above for $\beta$-decays. The $\alpha$-spectrum as
a function of time is given in the bottom panel of Figure
\ref{fig:netSpec}. The energy is fairly evenly distributed in the
range 5 MeV $< E_{\alpha} < 9$ MeV.

\subsubsection{Fission}

Spontaneous, neutron-induced, and \bt-delayed fission of heavy nuclei
($A \gtrsim 250$) contribute a few percent of the total \rp\ radioactive decay energy at
times $\lesssim 1$ day.  The mass distribution and energy spectra of
the fission fragments depend sensitively on the nuclear physics
models, and a thorough exploration of these parameters is beyond the
scope of this work. We can, however, estimate the final kinetic energy
of fission as equal to the repulsive Coulomb energy between the
daughter nuclei immediately after fission occurs:
\begin{equation}
E_{\rm K,tot} = E_{\rm Coul} = \frac{Z_1 Z_2 e^2}{r_0\left( A_1^{1/3} + A_2^{1/3}\right)},
\end{equation}
where $e$ is the elementary charge, $(A_1, Z_1)$ and $(A_2, Z_2)$ are the masses and atomic numbers of the daughter nuclei, and the nuclear radius is given by $r_0 A^{1/3}$. For deformed post-scission nuclei, $r_0 \simeq 1.8$ fm.

Fission favors the production of nuclei at or near the doubly-magic nucleus $(A, Z) = (132, 50)$. 
Assuming a typical parent isotope has mass and atomic numbers $A_p = 250$ and $Z_p = 100$, and that ($A_1, Z_1) = (132, 50)$, the fission daughters will have kinetic energies of order 100 MeV. We assume the fission fragment spectrum is flat, and ranges from 100--150 MeV. Given the limited role of fission at times later than 1 day, a more detailed treatment is unnecessary.

\section{Thermalization Physics}
\label{sec:thermRates}

In this section, we discuss the processes by which energetic decay products thermalize in the kilonova ejecta, and present energy loss rates for \bt-particles, \al-particles, and fission fragments.

\subsection{Gamma-rays}

Gamma-rays lose energy through photoionization and Compton
scattering.  
We calculated
the Compton opacity from the Klein-Nishina formula and the
photoionization opacity using the Photon Cross Section Database
(\textit{XCOM}; \citealt{NIST_XCOM}) published by the National
Institute of Standards and Technology (NIST).

\begin{figure}
\includegraphics[width=3.5 in]{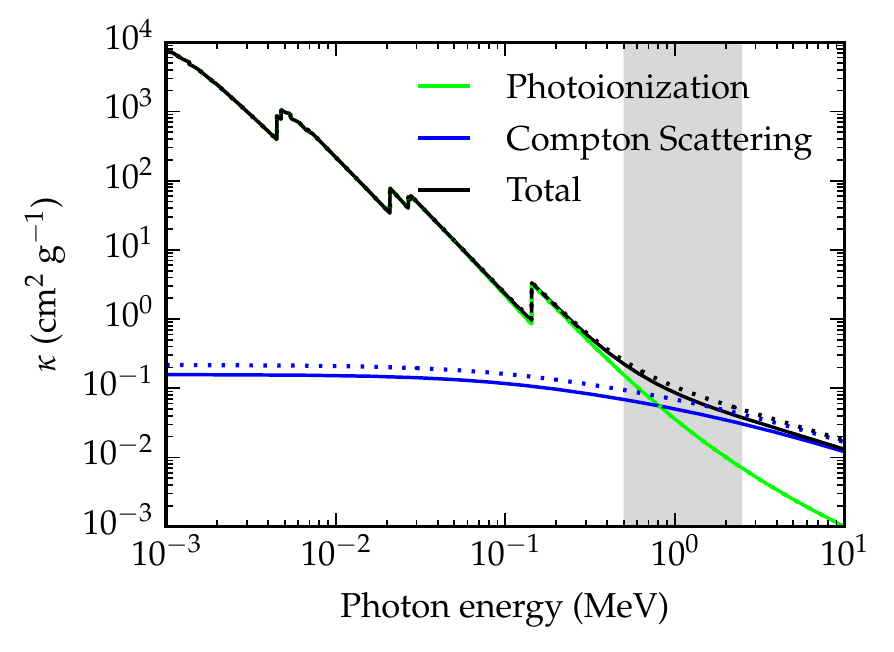}
\caption{The $\gamma$-ray opacity, $\kappa$, in the inner (solid lines) and outer (dotted lines) ejecta. Photoionization opacity is plotted in green, Compton opacity in blue, and total opacity in black. Differences between the inner and outer ejecta compositions have a negligible impact on $\kappa$. The gray bar indicates the energies at which most \g-rays are emitted.}
\label{fig:gammaOp}
\end{figure}

The total \g-ray opacity for our fiducial composition at $t=1$ day is shown in Figure~\ref{fig:gammaOp}.
The high-$Z$ elements produced in the \rp\ have higher ionization thresholds ($\sim 100$~keV) than do the metals in  typical astrophysical mixtures, so the photoionization cross-section in kilonovae dominates out to $\sim 1$~MeV, above which
Compton scattering takes over.
The opacity varies little between the inner and outer ejecta, and changes over time are minor, so we assume the \g-ray opacity to be constant.

Both photoionization and Compton scattering events produce a non-thermal electron, which loses energy by the physical processes described in the next section.

\subsection{Beta particles}
Suprathermal \bt-particles lose energy primarily through Coulomb interactions with free thermal electrons (plasma losses), and by exciting or ionizing bound atomic electrons. Bremsstrahlung (free-free) emission is important for very high-energy \bt-particles. While earlier studies of thermalization assumed plasma interactions dominated the energy loss, we find \bt-particles lose most of their energy to ionization and excitation.

In the limit that the \bt-particle energy far exceeds that of thermal electrons, the plasma energy loss per unit time is \citep{Huba_NRL}
\begin{align}
\dot{E}_\bt^{\rm pl} &=  7.7 \times 10^{-15} E_\bt^{-1/2} \\ 
&\times
\left( \frac{n_{\rm e}}{1 \text{ cm}^{-3}} \right) \lambda_{\rm ee} \left(1.0 - \frac{3.9}{7.7}\frac{T}{E_\bt} \right) 
\mbox{ MeV s}^{-1}
\end{align}
where $E_\bt$ is the \bt-particle's kinetic energy in MeV, $T$ is the ejecta temperature in MeV, $\lambda_{\rm ee} \sim 10$ is the Coulomb logarithm for electron-electron scattering, and $n_{\rm e}$ is the free electron number density.
Radioactive \bt-particles have $E_\bt \sim 1$ MeV whereas $k_{\rm B} T \sim 1$ eV in kilonova ejecta, so the assumption  that $E_{\bt} \gg k_{\rm B} T$ holds. 
We determine $n_{\rm e}$ assuming that all elements heavier than hydrogen are singly ionized, as expected for kilonova ejecta near peak brightness 
\citep{Kasen_2013_AS}.  The outer ejecta contains a substantial quantity of hydrogen, which we assume to be neutral given the low temperatures ($T \lesssim 5000$~K) expected in the ejecta periphery. 

We calculate energy losses due to ionization and excitation of atomic electrons using the well-established formula (\citealt{Heitler_1954_qntmRad,Berger_Seltzer_1964,Gould_1975,Blumenthal_Gould_1970}; see also \citet{Chan_Lingen_1993} and \citealt{Milne_1999_p+})
\begin{gather}
\begin{align}
\dot{E}_\bt^{\rm IE}
&=\frac{2 \pi r_{\rm e}^2 m_{\rm e} c^3 n_{\rm e,b}}{v_\bt/c} \\
&\times  \left\{ 2\ln\left(\frac{\displaystyle E_\bt}{\displaystyle \bar{I}}\right) + \ln\left(1 + \frac{\tau}{2}\right) + \left(1 - \frac{v_\beta^2}{c^2}\right)g(\tau) \right\}, \nonumber
\end{align}
\\ g(\tau) = 1 + \frac{\tau^2}{8} - (2 \tau+1)\ln 2,
\end{gather}
where 
$r_{\rm e}$ is the classical electron radius, 
$m_{\rm e}$ is the electron mass, 
$n_{\rm e,b}$ is the number density of bound electrons,
$v_{\beta}$ is the \bt-particle's speed,
and $\tau = E_\bt/m_{\rm e} c^2$. The quantity $\bar{I}$ is an average ionization and excitation potential which can be approximated for an element of atomic number $Z$ as
\citep{Segre_NucPart}
\begin{equation}
\bar{I} = 9.1Z\left(1 + \frac{1.9}{Z^{2/3}} \right) \text{ eV}.
\end{equation}
Following \citet{Chan_Lingen_1993}, we use averaged quantities for $n_{\rm e,b}$ and $\bar{I}$,
\begin{gather}
\langle n_{\rm e,b} \rangle = \frac{\rho}{m_{\rm u}} \left\langle \frac{Z}{A} \right\rangle,\\
\left\langle \ln \frac{\bar{I}}{\text{eV}}\right\rangle = \left\langle\frac{Z}{A}\right\rangle^{-1}\sum\limits_{\rm j} \left( \frac{A}{Z}\right)_{\rm j} X_{\rm j} \ln \left(\frac{\bar{I}_{\rm j}}{\text{eV}}\right),  \\
\text{where}~~ \left\langle \frac{Z}{A}\right\rangle = \sum\limits_{\rm j}\left(\frac{Z}{A}\right)_{\rm j}X_{\rm j},
\end{gather}
where $m_{\rm u}$ is the nuclear mass unit, 
$X_{\rm j}$ is the mass fraction of
element $j$, and the sum runs over all species in the composition. For
the inner (outer) ejecta, we find
$\langle \ln \bar{I}/\text{eV} \rangle = 6.4$ ($4.9$), and
$\langle Z/A \rangle = 0.4$ ($0.55$).

Plasma and ionization/excitation losses are the cumulative results of many distant interactions that individually transfer very little energy. The thermal and bound electrons energized by \bt-particles through these channels have very low kinetic energies, and thermalize rapidly.
Instead of tracking secondary electrons explicitly, we assume their kinetic energy is transferred directly to the thermal pool.

Bremsstrahlung (free-free) and synchrotron emission are other possible
means of \bt-particle energy loss. The rate of cooling due to
synchrotron emission in a magnetic field $B$ is 
\begin{equation}
\dot{E}_{\bt}^{\rm syn} = \frac{4}{9} r_{\rm e}^2 c\gamma^2 \left(\frac{v_{\beta}}{c}\right)^2 B^2,
\end{equation} 
where $\gamma$ is the \bt-particle's Lorentz factor.
Neglecting logarithmic terms (which only increase $\dot{E}_{\bt}^{\rm
  IE}$), and assuming $\langle Z/A\rangle$ = 0.4 and
$\gamma^{2}(v_{\bt}/c)^{3} \approx 10$, we estimate the ratio of
synchrotron to ionization/excitation losses as 
\begin{equation}
\frac{\dot{E}_{\bt}^{\rm syn}}{\dot{E}_{\bt}^{\rm IE}} \sim 1.6 \times 10^{-15} \left( \frac{B_{\rm d}}{3.7 \times 10^{-6} \text{ G}}\right)^2 M_5^{-1} v_2^{-1} t_{\rm d}^{-1},
\end{equation}
where $M_5 = M_{ej}/(5 \times 10^{-3} \Msun)$
and $B_{\rm d}$ is the magnetic field at 1 day. 
This is much less than unity for all parameters of interest, so we neglect synchrotron losses.

In contrast, Bremsstrahlung contributes, albeit modestly, to \bt\ energy loss for $E_\bt\gtrsim 1$ MeV. From \citet{Seltzer_Berger_Brem_1986},
\begin{equation}
\dot{E}_{\beta}^{\rm Brem} = n_{\rm i} v_{\bt} (E_\beta + m_{\rm e} c^2) Z^2 r_{\rm 0}^2 \alpha \phi_{\rm rad}
\end{equation}
where $n_i$ is the number density of the scattering species, $\alpha$ is the fine-structure constant, and $\phi_{\rm rad}$ are energy-dependent empirical fitting constants, also from \citet{Seltzer_Berger_Brem_1986}. We model Bremsstrahlung losses in the inner ejecta using characteristic values $Z = 60$ and $A = 144$,
similar to the average values of the inner composition.
For the outer ejecta, we use a two component composition, with $(Z,A)=(1,1)$ accounting for the high amount of hydrogen, and $(Z,A) = (55,133)$ representing elements with $Z > 1$.

Bremsstrahlung may produce high-energy photons that do not thermalize promptly in the ejecta, an effect of order $\lesssim 10\%$ at typical \bt-particle energies. Our treatment of Bremsstrahlung is discussed in more detail in \S \ref{subsec:partTrans}.

\begin{figure}
\includegraphics[width=3.25 in]{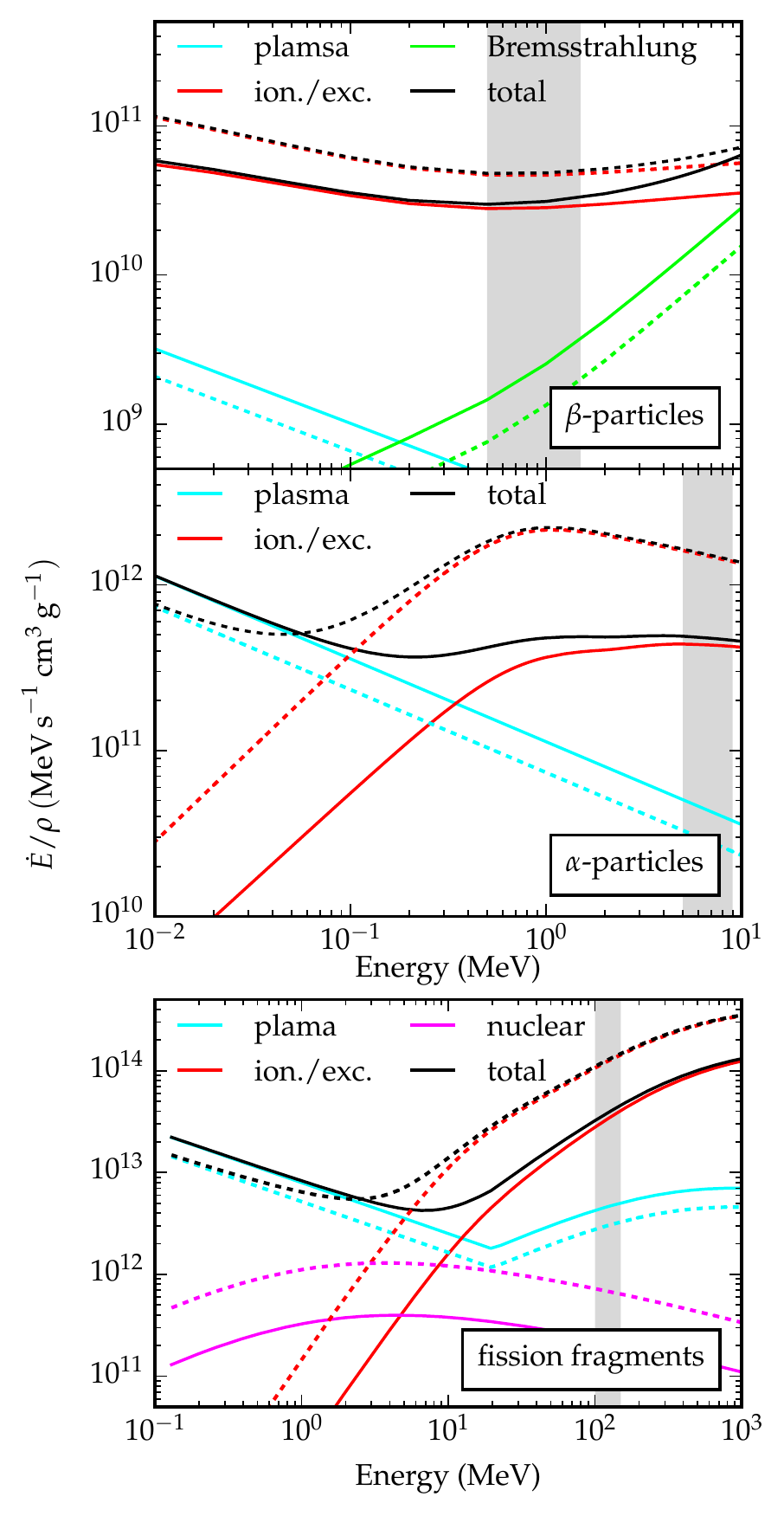}
\caption{\textit{All panels}: Energy loss rates in the inner (outer) ejecta are plotted in solid (dashed) lines. Gray bars indicate typical particle energies at emission. \textit{Top panel:} The total energy loss rate for fast \bt-particles, normalized to the mass density $\rho$. Thermalization rates are higher in the outer ejecta by a factor of a few. \textit{Middle panel:} The energy loss rates for \al-particles in simplified \rp\ mixtures standing in for the full inner and outer ejecta compositions (see Figure \ref{fig:rpSimpComp}), normalized to density. Alpha particle thermalization is a few to $\sim 10 \times$ more efficient in the outer ejecta. \textit{Bottom panel:} The energy loss rate for fission fragments, normalized to density and assuming most atoms in the ejecta are singly ionized. Thermalization is more efficient in the outer ejecta.}
\label{fig:dEdt_all}
\end{figure}

We plot the total energy loss rate in the inner and outer ejecta, normalized to mass density, in the top panel of Figure~\ref{fig:dEdt_all}. While the lower degree of ionization in the outer ejecta makes plasma and Bremsstrahlung losses less efficient, this is more than compensated for by enhanced ionization and excitation losses due to the greater number of bound electrons per nucleon, and to the lower average ionization potential. Overall, thermalization rates in the outer ejecta are higher by a factor of a few.

\subsection{Alpha-particles}
Suprathermal \al-particles thermalize by interacting with free and bound
electrons. Long-range interactions with ions and short-range
interactions with atomic nuclei do not significantly contribute to
\al-particle energy loss.  

Alpha particles scattering off of free, thermal electrons lose energy at a rate given by
\citet{Huba_NRL} for fast ions in a plasma, 
\begin{equation}
\begin{split}
\dot{E}_{\rm i}^{\rm pl} &=  1.7 \times 10^{-13} E_{\rm i}^{-1/2}\mu_{\rm i}^{1/2} Z_{\rm i}^2 \left( \frac{n_{\rm e}}{1 \text{ cm}^{-3}}\right) \lambda_{\rm ie} \\
&\times \left(2 - \frac{1.1\times 10^{-3}}{\mu_{\rm i}} - \frac{T}{E_{\rm i}} \right) \text{ MeV s}^{-1},
\end{split}
\label{eq:plasma_ie}
\end{equation}
where $E_{\rm i}$ is the ion's kinetic energy in MeV, $\mu_{\rm i}$ is the ion mass in $m_{\rm u}$, $Z_{\rm i}$ is the charge in units of the elementary charge, and $\lambda_{\rm ie} \sim 5 - 10$ is the Coulomb logarithm for ion-electron scattering. For \al-particles, $Z_{\rm i} = 2$ and $\mu_{\rm i} = 4$.

\begin{figure}
\includegraphics[width=3.5 in]{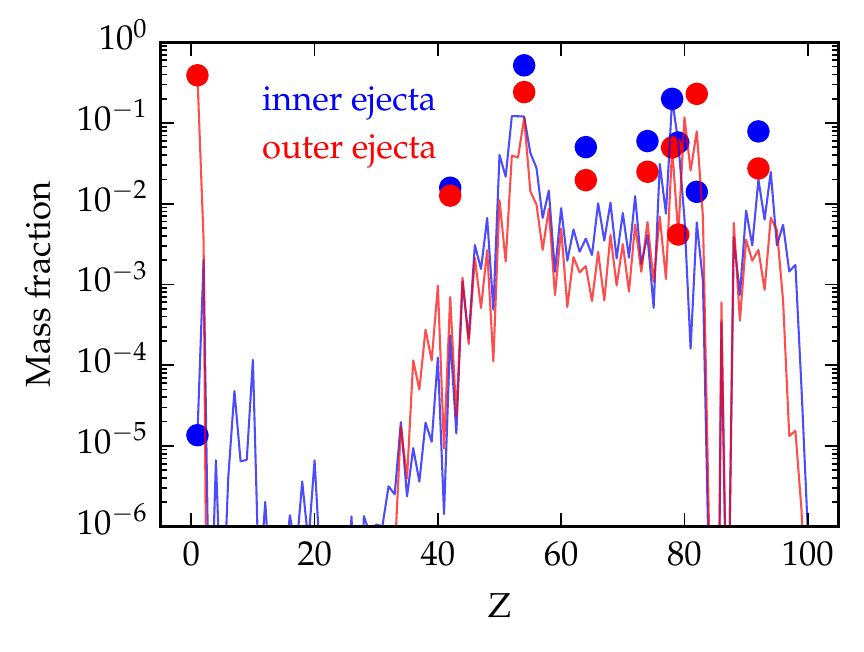}
\caption{The simplified composition (circles) used to calculate the electronic stopping for \al-particles and fission fragments. The full compositions, shown in solid lines, were mapped onto a composition of elements for which \al- and proton stopping data were available through NISTS's \textit{ASTAR} database.}
\label{fig:rpSimpComp}
\end{figure}

The rates of \al-particle energy loss due to interactions with bound electrons
are taken from  NIST's \textit{ASTAR} database \citep{NIST_*Star}.
Lacking \al-particle stopping data for all elements in our \rp\ mixture, we map the full inner and outer compositions onto a reduced set of elements for which \al-stopping powers are available (see Figure~\ref{fig:rpSimpComp}). The middle panel of Figure~\ref{fig:dEdt_all} shows the total \al-particle energy loss rates. Plasma losses dominate for $E_{\al} \lesssim 1$ MeV, while interactions with bound electrons are important at higher energies. The thermalization rate in the outer ejecta is greater than in the inner ejecta by up to an order of magnitude.

\subsection{Fission fragments}

Interactions with free and bound electrons, and with atomic nuclei all contribute to fission fragment thermalization.
The energy loss to thermal free electrons is described by Eq. \ref{eq:plasma_ie}, where  $Z_{\rm i }$ depends on the ionization state of the fission fragment and the length scale of the collision. For impact parameters greater than the size of the interacting particles, the relevant charge is the total charge carried by the fragment, $Z_{\rm ff,ion} = Z_{\rm nuc} - N_{\rm e^-,b}$, where $Z_{\rm nuc}$ 
is the fragment's atomic number and $N_{\rm e^-,b}$ is the number of bound electrons. At lower impact parameters, more of the nuclear charge is felt.

We calculate $Z_{\rm ff,ion}$ as a function of fission fragment energy $E_{\rm ff}$ using the formula of 
\citet{Schiwietz_Grande_2001_chargeState} for ion charge state in a gaseous medium. 
Since fragments with $Z_{\rm ff,ion} \gtrsim 7$ deflect thermal electrons at impact parameters greater than the fission fragment radius, we set $Z_{\rm i} \rightarrow \max\{Z_{\rm ff,ion}(E_{\rm ff}), 7\}$. 
Fission fragments can scatter off thermal \emph{ions} at much lower impact parameters, in which case the full nuclear charge is felt. The energy loss from these interactions is given by the nuclear stopping formula of \citet{Ziegler_1980}.

To model the stopping of heavy particles by bound electrons, we adopt the technique of \citet{Ziegler_1980}, in which the stopping power of a heavy particle in any material is proportional to the stopping power of a proton in the same material, with the constant of proportionality given by $Z_{\rm ff,ion}^2$. We calculate the stopping power for the same simplified composition used to model \al-particle energy loss, using proton stopping powers extracted from NIST's \textit{PSTAR} database \citep{NIST_*Star}.

The total energy loss rate for fission fragments is presented in the bottom panel of Figure \ref{fig:dEdt_all}. Interactions with bound electrons dominate the rate at high energies, while losses to free electrons become important at energies less than $\sim 10$ MeV. Thermalization rates in the outer composition are a factor of a few higher than in the inner composition.

\section{Analytic Results}
\label{sec:analytics}
Before moving to detailed numerical calculations of kilonova thermalization, we consider simple analytic estimates of the relevant timescales and time evolution. This work extends the analytic treatments proposed by 
\citet{Metzger_2010,Hotokezaka_2015_heating}.
Unless stated otherwise, our estimates describe thermalization in the ``inner'' composition, which typically makes up $\sim 90$\% or more of the ejected mass.

\subsection{Analytic estimates of thermalization timescales}
\label{subsec:anEst_Sum}
The net thermalization of the energy from the radioactive decay of \rp\ material depends on the relative importance of each decay channel and on how efficiently the decay products thermalize in the ejecta. 
Energy loss rates depend on the density of the medium, so thermalization is also a function of \mej\ and \vej. If we approximate the ejecta as a uniform density sphere of mass \mej\ and kinetic energy $E_{\rm k} = \mej\vej^2/2$, the density is 
\begin{equation}
\rho(t) \approx 7.9 \times 10^{-15} M_{5} \; v_{2}^{-3} \; t_{\rm d}^{-3} \text{ g cm}^{-3},
\end{equation}
where again, $M_5 = \mej/5.0 \times 10^{-3} M_{\odot}$ and $v_2 = \vej/0.2c$.
Thermalization becomes inefficient at a time, $t_{\rm ineff}$, when
the timescale for a particle to thermalize becomes similar to the
ejecta expansion timescale, $t_{\rm exp}$. The inefficiency time can
be compared to the peak of the kilonova light curve,
\begin{equation}
t_{\rm peak} \sim \left(C \frac{M_{\rm ej} \kappa}{v_{\rm ej} c}\right)^{1/2} \simeq 4.3 \; M_{5}^{1/2} \; v_2^{-1/2} \: \text{days},
\end{equation}
where $\kappa$ is the opacity for optical/infrared light (we take $\kappa = 10 \text{ cm$^2$ g}^{-1}$, appropriate for an \rp\ medium) and $C = 0.32$ is a scaling factor we estimate from kilonova radiation transport simulations (e.g. Barnes \& Kasen 2013). If $t_{\rm ineff} < t_{\rm peak}$, thermalization will impact the kilonova light curve significantly.

\vspace{\baselineskip}
\noindent $\boldsymbol{\gamma}$\textbf{-rays:} Gamma rays stop thermalizing efficiently when they can escape the ejecta without undergoing any scatters or absorptions. This occurs when the optical depth $\tau \approx \rho \kappa_\g R_{\rm ej}$ falls below unity. 
For \g-rays with  energies $E_{\g} \gtrsim 1$ MeV, the relevant opacity is the Compton opacity, $\kappa_{\rm C} \approx 5 \times 10^{-2}$ cm$^2$ g$^{-1}$ while the photoionization opacity, $\kappa_{\rm PI} \gtrsim 1$ cm$^2$ g$^{-1}$, dominates for lower-energy photons. The ejecta becomes transparent ($\tau < 1$) to \g-rays at a time
\begin{align}
t_{\rm ineff} \approx& \begin{cases}
0.5 \: M_5^{1/2} \: v_2^{-1} \mbox{ days} &\mbox{for }  E_{\g} \gtrsim 1 \mbox{ MeV} \\
2.3 \: M_5^{1/2} \: v_2^{-1} \mbox{ days} &\mbox{for } E_{\g} \lesssim 1 \mbox{ MeV}
\end{cases}
\end{align}
In both cases, inefficiency sets in before the kilonova light curve peaks,
\begin{subnumcases}{\frac{t_{\rm ineff}}{t_{\rm peak}} \simeq }
0.12 \: v_2^{-1/2} \hfill &\mbox{ } $E_{\g} \gtrsim 1 \mbox{ MeV}$ \label{eq:titp_g1} \\ 
0.5 \: v_2^{-1/2} \hfill &\mbox{ } $E_{\g} \lesssim 1 \mbox{ MeV.}$ \label{eq:titp_g2}
\end{subnumcases}

\vspace{\baselineskip}
\noindent $\boldsymbol{\beta}$ \textbf{-particles:}  The energy loss rate for \bt-particles, modulo mass density, has a fairly constant value $\dot{E}_{\bt} \simeq 4 \times 10^{10} \rho \text{ MeV s}^{-1} $ over a broad range of energies (see Fig. \ref{fig:dEdt_all}). The thermalization time for \bt-particles is
\begin{eqnarray}
t_{\rm th} &\approx & \frac{E_{\bt,0}}{\dot{E}_{\bt,0}} = \frac{E_{\bt,0}}{ 4 \times 10^{10} \; \rho \text{ MeV s}^{-1}} \nonumber \\
&=& 0.02 \left(\frac{E_{\bt,0}}{0.5 \text{ MeV}}\right) \; M_5^{-1} \; v_2^{3} \; t_{\rm d}^3  \text{ days},
\end{eqnarray}
where $E_{\bt,0}$ is the initial \bt-particle energy.

Beta particles trapped in the ejecta fail to efficiently thermalize
when $t_{\rm th} \gtrsim t_{\rm exp}$,  
which occurs at
\begin{equation}
t_{\rm ineff}  \approx   7.4 \: \left(\frac{E_{\bt,0}}{0.5 \text{ MeV}}\right)^{-1/2} \; M_5^{1/2} \; v_2^{-3/2} \text{ days}. 
\end{equation}
For a typical initial energy, $t_{\rm ineff}$ is comparable to the rise time of the light curve,
\begin{align}
\frac{t_{\rm ineff}}{t_{\rm peak}} \approx 1.7 \left(\frac{E_{\bt,0}}{0.5 \text{ MeV}}\right)^{-1/2} \; v_2^{-1}. \label{eq:titp_beta}
\end{align}

If the magnetic field is radial or only slightly tangled, \bt-particles can escape the ejecta before they thermalize, and escape will significantly reduce the thermalization efficiency. The escape time is 
\begin{equation}
t_{\rm esc} \simeq \frac{R_{\rm ej}(t)}{\lambda v_{\bt,\parallel}}, 
\end{equation}
where $\lambda R_{ej}$ is the coherence length of the magnetic field, $v_{\bt,\parallel}$ is the component of the \bt-particle velocity parallel to the field lines, and we have modeled the \bt's motion in a random field as a random walk of step size $\lambda R_{\rm ej}$. 
For a \bt-particle with $E_{\rm \bt, 0} = 0.5$ MeV and pitch angle 1 ($v_{\bt,\parallel} = v_{\beta}$), $t_{\rm esc}$ is less than $t_{\rm th}$ when
\begin{equation}
t \gtrsim \frac{3.5 \; M_5^{1/2} \; v_2^{-1}}{\lambda^{1/2}}~~{\rm days}.
\end{equation}
For radial fields ($\lambda = 1$), this is less than $t_{\rm peak}$, so escape is important for \bt-particle thermalization.
In contrast, for disordered fields there is a degree of randomness above which \bt-particle escape cannot significantly impact the light curve. This limit is defined by the condition $t_{\rm th}(t_{\rm peak}) < t_{\rm esc}(t_{\rm peak})$. Again considering a 0.5
MeV \bt-particle, we find
\begin{equation}
t_{\rm th}({t_{\rm peak}}) < t_{\rm esc}(t_{\rm peak}) \rightarrow \lambda \lesssim 0.8 v_2^{-1}. \\
\end{equation}
Thus, high-energy \bt-particles are effectively trapped by even a slightly tangled magnetic field.

\vspace{\baselineskip}
\noindent $\boldsymbol{\alpha}$\textbf{-particles and fission fragments:} 
Fission fragments and \al-particles are emitted with greater energies than \bt-particles ($E_{\al,0} \simeq 6$ MeV; $E_{\rm ff,0} \simeq 100$ MeV), but have higher energy loss rates ($\dot{E}_{\al}(E_{\al,0}) \sim 5 \times 10^{11}\rho$ MeV s$^{-1}$; $\dot{E}_{\rm ff}(E_{\rm ff,0}) \sim 5 \times 10^{13} \rho$ MeV s$^{-1}$.) The efficiency of \al-particle thermalization is similar to that of \bt\ particles, while fission fragments thermalize efficiently out to very late times:
\begin{numcases}{\frac{t_{\rm ineff}}{t_{\rm peak}} \simeq }
1.8 \: \left(\frac{E_{\al,0}}{6 \text{ MeV}}\right)^{-1/2} v_2^{-1} \hfill &\mbox{ \al-particles} \nonumber \\
\label{eq:titp_a} \\
3.9 \: \left(\frac{E_{\rm ff,0}}{125 \text{ MeV}}\right)^{-1/2} v_2^{-1} \hfill &\mbox{ fiss. fragments.} \nonumber \\
\label{eq:titp_ff}
\end{numcases}

Unlike \bt-partices, both \al's and fission fragments have velocities much lower than $v_{\rm ej}$, and so in general cannot escape the ejecta. However, because
these particles are
propagating through a steep velocity gradient, 
their speed relative to
the background gas continually decreases.  This reduces the kinetic energy
of the particles as measured in the co-moving frame.  Because the particles
have a spiraling motion about magnetic field lines, their motion is
never completely frozen out in the fluid frame.  Still, these ``frame-to-frame''
effects can reduce thermalization by $\lesssim 15\%$.

\subsection{Summary of thermalization timescales}
\begin{figure}
\includegraphics[width = 3.5 in]{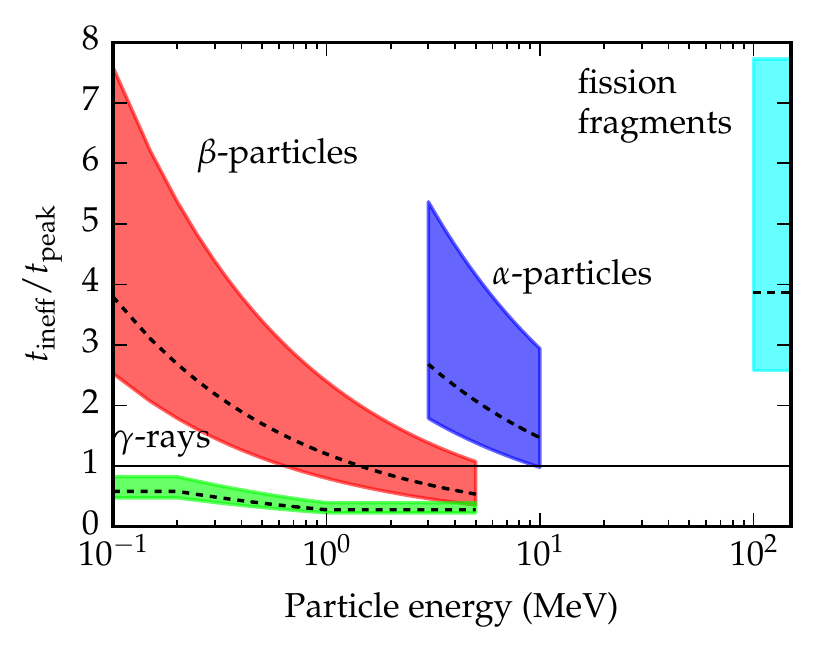}
\caption{The ratio $t_{\rm ineff}/t_{\rm peak}$ for all particles, for \vej\ in the range $0.1c - 0.3c$. Fission fragments, and to a lesser extent \al-particles and low-energy \bt-particles, thermalize efficiently out to late times. Higher energy \bt's and \g-rays are expected to become inefficient on kilonova timescales. The width of the curves is due to the range of $v_{\rm ej}$ considered, since $t_{\rm ineff}/t_{\rm peak}$ varies inversely with \vej. Curves for the fiducial velocity \vej = $0.2c$ are over-plotted in dotted black lines.}
\label{fig:anEstSum}
\end{figure}

While low-energy \bt-particles, \al-particles, and especially fission fragments typically thermalize efficiently at $t = t_{\rm peak}$, the thermalization at peak of high-energy \bt-particles and \g-rays is not robust. 
Figure~\ref{fig:anEstSum} plots the ratio of thermalization time to light curve peak for all particles as a function of initial energy for a range of $v_{\rm ej}$. 
For \al- and \bt-particles, we calculated $t_{\rm ineff}/t_{\rm peak}$ from Eq.s~\ref{eq:titp_a} and \ref{eq:titp_beta}. The \g-ray curve was calculated from Eq.~\ref{eq:titp_g1} for $E_{\g} \leq 200$ keV, \ref{eq:titp_g2} for $E_{\g} \geq 1$ MeV, and a simple linear interpolation for intermediate $E_{\g}$. For fission fragments, we modified Eq.~\ref{eq:titp_ff} slightly to account for the positive slope of $\dot{E}_{\rm ff}$ in the range $E_{\rm ff} = 100 - 150$ MeV. This renders $\dot{E}_{\rm ff}$ approximately constant, so the fission fragment curve is essentially flat. 

\subsection{Analytic thermalization model}
\label{subsec:anMod}

We develop an analytic expression for time-dependent thermalization
efficiencies of massive particles under the following
assumptions: first, that the radioactive energy generation rate evolves as
$t^{-\al}$ with $\al = 1.0$ (close to the expected values
$\al = 1.1-1.4$); second, that the density in the
ejecta is spatially uniform; third, that energy loss rates are
independent of particle energy, and depend only on $\rho$; and fourth,
that all particles of a given type are emitted at a single energy
$E_0$. Despite these simplifications, we find our model agrees fairly
well with the detailed numerical calculations to be presented in
\S\ref{sec:results}.

The thermalization efficiency is defined as the ratio of energy
emitted by radioactive processes to energy absorbed by the ejecta at
any time $t$, 
\begin{align}
f(t) = \frac{\dot{E}_{\rm th}(t)}{\dot{E}_{\rm rad}(t)}.
\end{align}
We approximate the radioactive energy generation rate  by
$\dot{E}_{\rm rad} = \dot{\epsilon}_0(t_0/t)$ with $\dot{\epsilon_0} =
10^{11} M_{\rm ej}$ ergs s$^{-1}$ and $t_0 = 1$ day. Assuming charged
particle thermalization depends only on mass density (which declines
like $t^{-3}$ in a homologous flow) the energy loss is 
\begin{equation}
\dot{E}_{\rm part}(t) =  \psi \rho_0 \left(\frac{t}{t_0}\right)^{-3},
\end{equation}
where $\rho_0$ is the density at $t_0$, and $\psi$ is a scaling factor
such that $\psi\rho_0 = \dot{E}_{\rm part}(t_0)$, which will be unique
to each particle type. The  rate at which energy is thermalized,
$\dot{E}_{\rm th}(t)$, is given by the number of live particles $N$
multiplied by the rate at which they lose energy, 
\begin{align}
\dot{E}_{\rm th}(t) = N(t) \times \psi \rho_0 \left(\frac{t}{t_0}\right)^{-3}.
\end{align}
At any time $t$, the oldest live particle originates from an earlier time $t_{\rm i}$, defined by 
\begin{align}
E_{\rm part}(t) &= E_0 - \int\limits_{t_{\rm i}}^t \psi \rho_0 \left(\frac{t'}{t_0}\right)^{-3} \dd{t'} = 0, \\
\intertext{ which is satisfied by }
t_{\rm i} &= \left(\frac{\psi \rho_0 t_0^3 t^2}{2 E_0 t^2 + \psi \rho_0 t_0^3}\right)^{1/2}.
\end{align}
The number of live particles at time $t$ is then
\begin{equation}
N(t) = \frac{\dot{\epsilon}_0 t_0}{2 E_0} \ln\left[1 + 2\left(\frac{t}{t_{\rm ineff}}\right)^2\right]
\end{equation}
where $t_{\rm ineff}$ is the inefficiency timescale defined in the previous section. 

It is now straightforward to calculate the ratio $f_{\rm p}$ of thermalized to emitted energy for a massive particle of type $p$,
\begin{equation}
f_{\rm p}(t) = \frac{\dot{E}_{\rm th}}{\dot{E}_{\rm rad}} = \frac{\ln\left[1 + 2\left(\frac{t}{t_{\rm ineff,p}}\right)^2\right]}{2\left(\frac{t}{t_{\rm ineff,p}}\right)^2}. 
\label{eq:ftAn}
\end{equation}
Eq.~\ref{eq:ftAn} can be used to estimate the thermalization efficiencies of massive particles, where the relevant timescales $t_{\rm ineff,p}$ are given by Eq.s~\ref{eq:titp_beta} (\bt-particles), \ref{eq:titp_a} (\al-particles), and \ref{eq:titp_ff} (fission fragments).

For \g-rays, the thermalization efficiency is approximately equal to the interaction probability: $f_{\g}(t) \approx 1 - e^{-\tau}.$
We can estimate the optical depth $\tau \approx \rho \kappa_{\g} R_{\rm ej}$ using $\bar{\kappa}_\g$, the \g-ray opacity averaged over the emission spectrum. Optical depth is related to $t_{\rm ineff,\g}$ by
\begin{gather}
\left(\frac{t_{\rm ineff,\g}}{t_0}\right)^{2} = \rho_0 \bar{\kappa}_\g R_0 = \tau_0 \nonumber \\
\rightarrow \tau(t) = \tau_0\left(\frac{t}{t_0}\right)^{-2} = \left( \frac{t_{\rm ineff,\g}}{t}\right)^2 \nonumber,
\intertext{so}
f_\g(t) =  1 - \exp\left[-\left(\frac{t}{t_{\rm ineff,\g}}\right)^{-2}\right] \label{eq:ftg_an}
\end{gather}

\begin{figure}
\includegraphics[width=3.5 in]{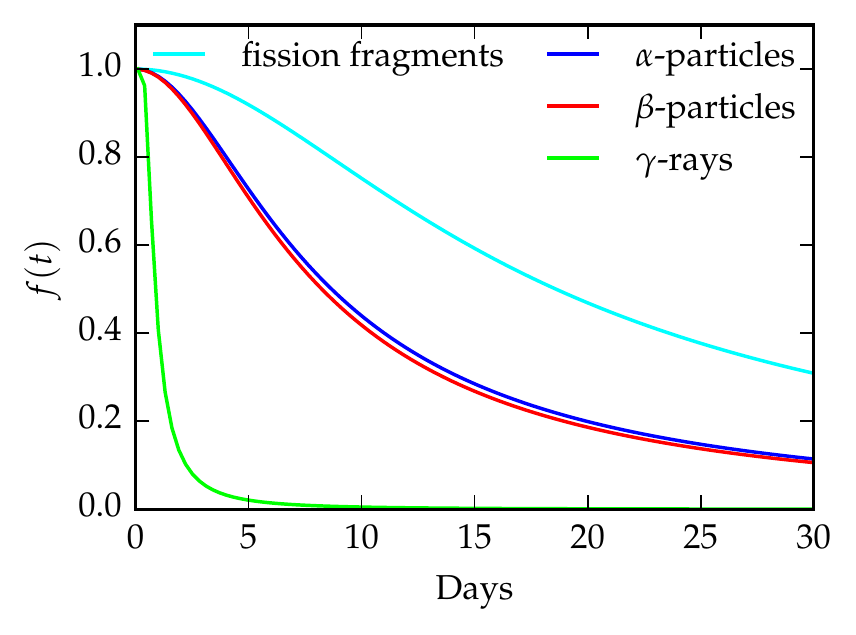}
\caption{Analytic thermalization efficiencies, calculated with Eq.s \ref{eq:ftAn} and \ref{eq:ftg_an}. We use $t_0 = 1$ day, and $\rho_0 = 7.9 \times 10^{-15}$ cm$^{-3}$, corresponding to a uniform density ejecta with the same mass and energy as our fiducial model. For \al's, \bt's, and fission fragments we take $E_0 = 6, 1,$ and $125$ MeV, respectively.}
\label{fig:anSimComp}
\end{figure}

Figure \ref{fig:anSimComp} shows our analytic thermalization functions for $\mej = 5 \times 10^{-3} M_{\odot}$, and $\vej = 0.2c$, using the expressions for $t_{\rm ineff}$ derived in \S \ref{sec:thermRates}. For massive particles, we used $E_{\bt,0} = 0.5$ MeV, $E_{\alpha,0} = 6$ MeV, and $E_{\rm ff,0} = 125$ MeV. For \g-rays, we take $\bar{\kappa} = 0.1$ cm$^2$ g$^{-1}$, which gives $t_{\rm ineff,\g} \approx 1.4$ days.

As we will see in \S \ref{sec:results}, the approximate analytic expressions Eq.s \ref{eq:ftAn} and \ref{eq:ftg_an} agree fairly well with our numerical results.

\section{Numerical Results}
\label{sec:results}

In this section, we present numerical calculations of thermalization efficiencies
as determined by modeling the 3-dimensional
transport of \g-rays, fission fragments, and \al- and \bt-particles in
a magnetized expanding medium.  
Our calculations used the time-evolving emission spectra introduced in \S \ref{subsec:spec}, accounted for the time-dependent partition of radioactive energy among different decay products, and incorporated the detailed, energy-dependent energy loss rates derived in \S \ref{sec:thermRates}. The flux tube approximation was used to model charged particle transport, allowing us to explore the sensitivity of our results to the architecture of the ejecta's magnetic field.
Additional details of our transport method
are given in the Appendix.

\subsection{Thermalization efficiencies}

\begin{figure}
\includegraphics[width=3.5 in]{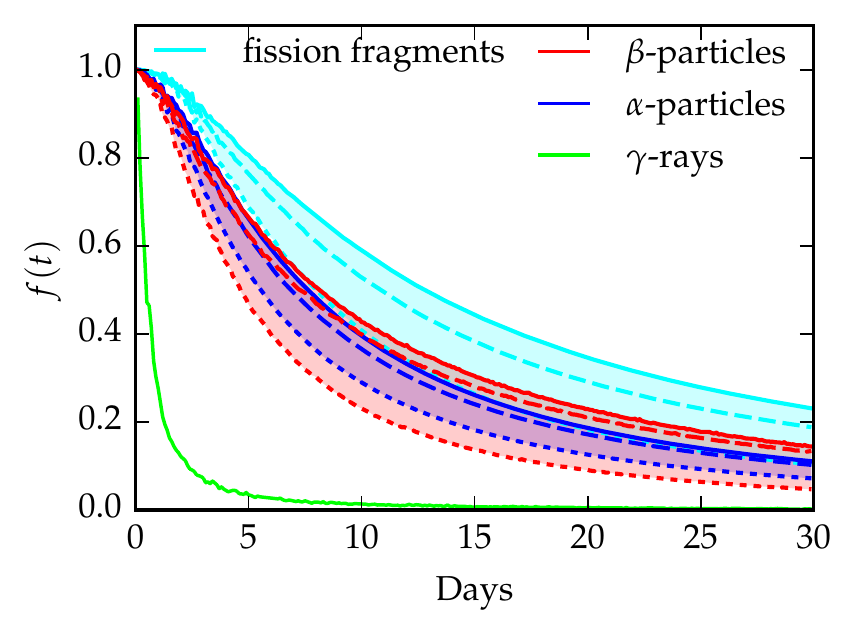}
\caption{Thermalization efficiencies $f(t)$ for all particles in an ejecta with $M_{\rm ej} = 5\times 10^{-3} \Msun$, and $v_{\rm ej} = 0.2c$ (our fiducial model). Fission fragments thermalize most efficiently, followed by \al-particles, \bt-particles, and \g-rays. For charged particles we plot $f(t)$ for radial (dotted lines), toroidal (solid lines), and moderately tangled ($\lambda = 0.25$; dashed lines) magnetic fields. Toroidal fields thermalize most efficiently, followed by random, then radial fields.}
\label{fig:ft_PartFid}
\end{figure}

Figure \ref{fig:ft_PartFid} presents the numerically calculated thermalization efficiency, $f(t)$, of all particles for the fiducial ejecta model (\mej = $5\times 10^{-3} \Msun$ and \vej = $0.2c$.) Fission fragments thermalize most efficiently, having $f(t) \gtrsim 0.5$ out to $t \sim 15$ days. Alpha- and \bt-particle thermalization is slightly lower, reaching $f(t)= 0.5$ around a week post-merger, while $f(t)$ for \g-rays is much lower, falling below 0.5 by $t\sim 1$ day. 

For massive particles, we show $f(t)$ for radial (dotted lines), toroidal (solid lines), and lightly tangled ($\lambda = 0.25$; dashed lines) magnetic field geometries. 
The magnetic field configuration
affects thermalization in three ways:
\begin{enumerate}
\item \textbf{Diffusion:} Radial or lightly tangled fields allow particles to diffuse outward into regions of lower density, and lead to lower $f(t)$. 
\item \textbf{Escape:} Radial fields that allow charged particles to escape before they have completely thermalized will lower $f(t)$. This is most important for \bt-particles, which move faster than the ejecta.
\item \textbf{Frame-to-frame effects:} Particles in a homologous flow
  lose energy, as measured in the co-moving frame (cmf), as they move through the ejecta. These frame-to-frame losses reduce the amount
  of kinetic energy a particle has to thermalize, and therefore reduce
  $f(t)$. Radial fields and lightly tangled fields, which allow particles to move fairly freely through the ejecta, facilitate frame-to-frame effects.
  These losses are most important for \al-particles and fission
  fragments, which have velocities of order \vej, and thus have
  substantially different cmf energies in different regions of the
  ejecta.
\end{enumerate}

In light of the above, it is not surprising that toroidal fields maximize $f(t)$; toroidal fields hold particles at one position in velocity space, preventing diffusion, escape, and frame-to-frame losses. Radial fields, in contrast, enhance all
three of these effects and hence minimize $f(t)$. 
Thermalization in random fields 
falls between these two extremes. This behavior holds for all ejecta models studied.

\begin{figure}
\includegraphics[width = 3.5 in]{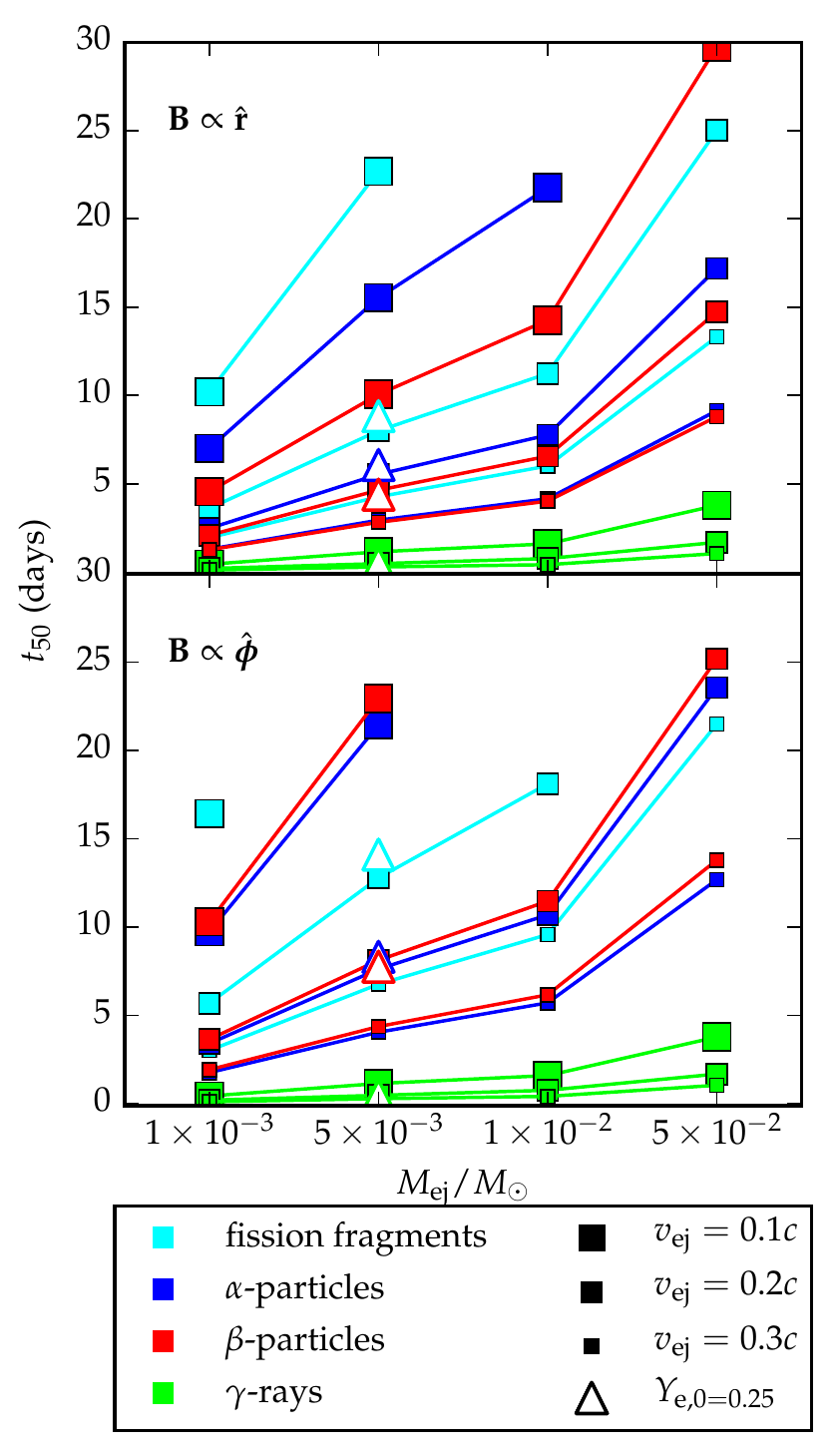}
\caption{ The time at which $f(t)$ drops below 50\% ($t_{50}$) for all
  particles, for all \mej\ and \vej\ considered. Results for a radial
  (toroidal) magnetic field are shown in the top (bottom)
  panel. Thermalization increases with mass and decreases with
  velocity. Fission fragments thermalize most efficiently, followed by
  \al-particles and \bt-particles, and finally \g-rays. Toroidal
  fields result in more robust thermalization of all massive
  particles.}
\label{fig:t50}
\end{figure}

While the trends shown in Figure~\ref{fig:ft_PartFid}---i.e., that
$f_{\rm ff}(t) > f_{\al}(t) \approx f_\bt(t) > f_\g(t)$---are
consistent across ejecta models, the values of $f(t)$ can
vary significantly with \mej\ and \vej.  Figure \ref{fig:t50}
illustrates the variance and clarifies the dependence of $f(t)$ on the
ejecta parameters. For each point (\mej, \vej) in parameter space, and
for each particle type, we plot $t_{50}$---the time at which $f(t)$
drops to 50\%. Cases in which $f(t = 30 \: {\rm days}) > 50\%$ are
omitted from Figure \ref{fig:t50}.) To show how sensitive
thermalization is to magnetic fields, we include results for radial
(top panel) and toroidal (bottom panel) field geometries.

The thermalization of all particles increases with \mej\ and decreases
with \vej. The changes in efficiency are especially dramatic for
massive particles. For the heaviest ejecta mass considered
($\mej = 5 \times 10^{-2} \Msun$), massive particles thermalize
efficiently out to late times regardless of $\vej$. The thermalization
of \g-ray energy is low for all models tested.

Though the results shown are for the two-component composition described above, we find that the higher energy loss rates in the outer ejecta have only a small impact on thermalization. For radial fields, where the effect is most pronounced, the two-zone model results in an increase in total thermalization of $< 5$\% relative to a one-zone model that assumes the entire ejecta has the ``inner'' ejecta composition. The $f(t)$ we calculate should be fairly insensitive to the exact division between inner and outer ejecta.

\subsubsection{Effect of aspherical ejecta}
The ejecta from a CO merger is likely to be aspherical, particularly in the case of NSBH merger, where most of the ejected mass is confined to the equatorial plane \citep[e.g.,][]{Hotokezaka_2013_massEj}.  To estimate the effect on thermalization, we compare a spherical model to an oblate one, where both models have \mej = $5\times 10^{-3}\Msun$ and \vej = $0.2c$, radial magnetic fields, and a broken power law density profile with $(\delta, n) = (1,10)$. For the oblate geometry, the density is a function of $\tilde{v}$, where
\begin{equation*}
\tilde{v} = v\sqrt{a^{-2/3}\sin^2\theta + a^{4/3}\cos^2\theta}
\end{equation*}
is chosen so that isodensity contours are oblate spheroids of aspect ratio $a$. 

\begin{figure}
\includegraphics[width = 3.5 in]{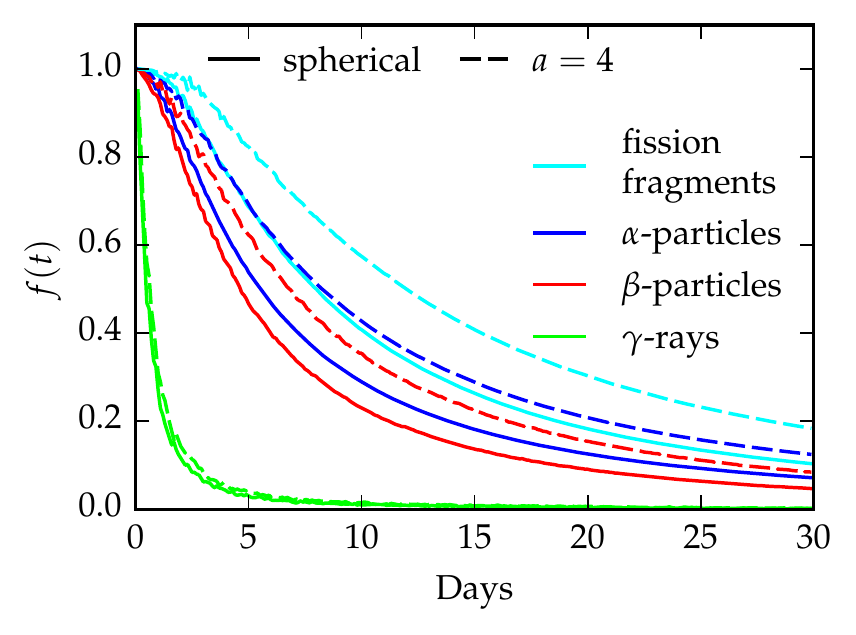}
\caption{Thermalization efficiencies for oblate ejecta with aspect ratio $a=4$, compared to the standard spherical geometry, for the fiducial mass and velocity and radial magnetic fields. Thermalization increases with increasing asymmetry. We found similar increases for random and toroidal fields.}
\label{fig:sphVOb}
\end{figure}

Figure \ref{fig:sphVOb} compares the $f(t)$ for the oblate  and spherical cases, and shows
that massive particle thermalization increases with increasing asphericity. For an aspect ratio $a=4$, the $f(t)$ for \al's, \bt's, and fission fragments increase by a factor of $\sim 1.5$ relative to spherical ejecta. Gamma-ray thermalization is higher for the oblate geometry, but only slightly. 
The higher $f(t)$ are due to the higher density of the oblate ejecta, which more than compensates for the increased ease of escape in directions perpendicular to the equatorial plane. 
Figure~\ref{fig:sphVOb} shows $f(t)$ only for radial magnetic fields, but we found similar increases for random and toroidal fields.

\subsection{Total heating efficiency}
\begin{figure}
\includegraphics[width = 3.5 in]{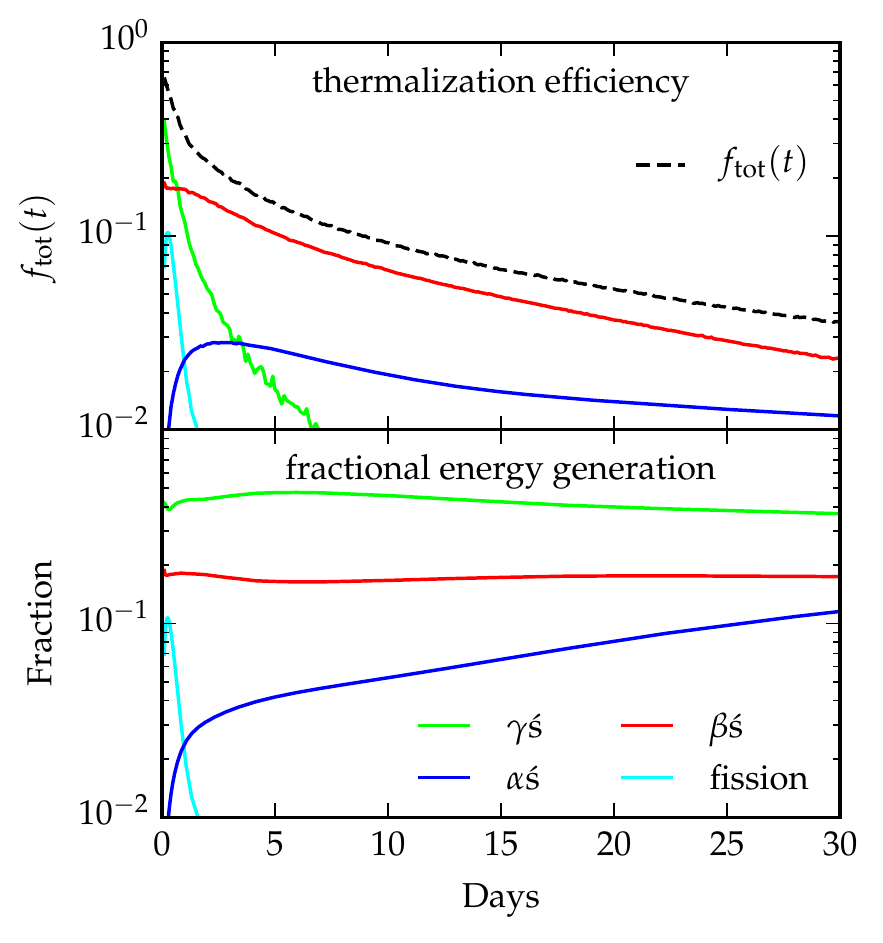}
\caption{\textit{Bottom panel:} The fractional energy generation associated with
  each type of particle, from \rp\ simulations using the FRDM mass
  model. The division of \bt-decay energy among \bt-particles,
  \g-rays, and neutrinos was calculated for our representative SPH
  trajectory with $Y_{\rm e, 0} = 0.04$. \textit{Top panel:} The fractions from the bottom panel, convolved with $f(t)$ for
  each particle, for the fiducial model with random magnetic
  fields. The total thermalization efficiency, $f_{\rm tot}$, plotted
  as a dashed black line, is the sum of the particle-specific
  curves. Beta- and \al-particles supply most of the thermalized
  energy.}
\label{fig:ftTot_FRDM}
\end{figure}

To study the net heating efficiency, we convolve $f(t)$
for each decay product with the fraction that particle contributes to  the total  energy generation.  
The bottom panel of Figure~\ref{fig:ftTot_FRDM} shows how 
the \rp\ decay

energy is divided among different particles, while the top panel shows the energy \emph{thermalized} by each particle
type, as a fraction of the total energy emitted across all decay channels. The $f(t)$ represented in Figure \ref{fig:ftTot_FRDM} are for a fiducial ejecta model with moderately tangled ($\lambda = 0.25$) magnetic fields. The total thermalization efficiency, which is simply a sum over particle types, is plotted in black. While \g's, \al's, \bt's, and fission fragments all have $f(t) \approx 1$ at very early times, the initial \emph{total} thermalization efficiency is less than one because a significant fraction of the \bt-decay energy is lost to neutrinos. 

The net thermalization efficiency, in this model, drops below 0.5 by $t=1$ day, and  below 0.1 by $t \sim 10$ days.
While \bt-particles and \g-rays dominate the energy production at all times, \g-rays thermalize inefficiently, and supply very little heating after $t \sim 1$ day. While \al-decay produces less than $\sim 10$\% percent of the total energy,  the $\alpha$-particles thermalize fairly efficiently, and so
 contribute a significant fraction of the total thermalized energy.

\begin{figure}
\includegraphics[width = 3.5 in]{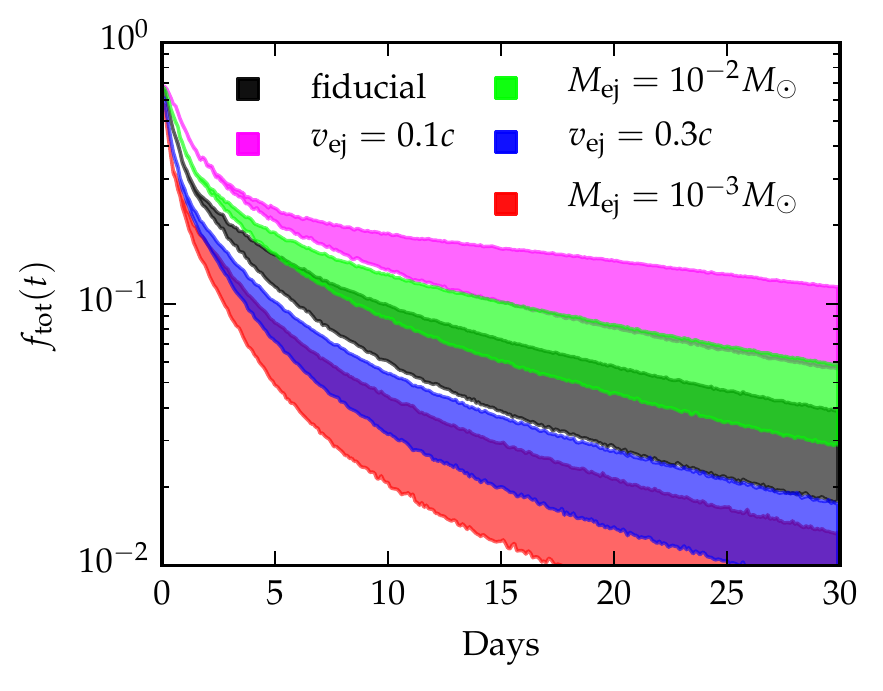}
\caption{Total thermalization efficiencies for different ejecta models
  (\mej, \vej) using FRDM energy generation rates. The fiducial model
  is plotted in black. Other curves differ from the fiducial model in
  \mej\ or \vej\ only. The width of the curves reflects the variation in
  $f(t)$ for different magnetic field configurations; the curves are
  bounded on top by \ftot\ for a toroidal field and on bottom by
  $\ftot$ for a radial field configuration. } 
\label{fig:ftTot_MV}
\end{figure}

The total heating efficiency has the expected dependence on the ejecta
parameters: greater masses and lower velocities lead to higher \ftot,
as shown in Figure \ref{fig:ftTot_MV}.  Thermalization for the low-mass and high-velocity models falls below 0.5 within a few days, and
below 0.2 by $5-7$ days. The high-mass and low-velocity models
thermalize much more efficiently, sustaining $f_{\rm tot}(t) > 0.5$
out to $t \lesssim 1$ week, and not falling below
$f_{\rm tot}(t) = 0.2$ until $t \sim 15 - 20$ days.  There is also
variation within each model (up to a factor of $\sim 2$) due to
uncertainties in the magnetic field.

\subsubsection{Dependence on nuclear physics}
\label{subsec:nucPhys}

The radioactive energy generation---and therefore the thermalization---depends
on \rp\ yields, which in turn are sensitive to variations in nuclear physics models and astrophysical conditions. To explore
this effect, we consider \rp\ yields computed for different mass
models, and for different initial $Y_{\rm e}$ of the ejected matter.
 
The yields differ primarily in the amount of translead nuclei synthesized relative to lighter \rp\ elements. 
\cite{MendozaTemis_etal_rProcess}
have shown that the production of translead nuclei is sensitive
to nuclear physics inputs, in particular to neutron
separation energies near $N=130$.
As discussed in \S\ref{subsec:comp}, the production of translead nuclei also depends on initial electron fraction, decreasing as $Y_{\rm e,0}$ increases.  

\emph{R}-process yields could impact thermalization in two ways. First, different yields have different abundance-averaged compositional properties, and could give rise to different thermalization rates.
Second, because nuclei heavier than lead decay mainly by fission and \al-emission, while lighter nuclei undergo \bt-decay, the amount of translead material will alter the relative importance of \al- and \bt-decay. Since all \al-decay energy is transferred to energetic \al-particles, which thermalize efficiently, while $\gtrsim 70$\% of \bt-decay energy goes to \g-rays and neutrinos, which do not, enhanced \al-decay may increase thermalization.   Based on these arguments,
we expect that differences in the amounts of translead nuclei
will result in different $f_{\rm tot}(t)$, and therefore, differences in predicted kilonova light curves.

To explore the strength of these effects, we compare the thermalization efficiency for three different compositions:
the reference \rp\ yields (based on the FRDM mass model); yields for the DZ31 mass model, which predicts increased production of
translead nuclei (see Figure~\ref{fig:comp}); and yields from a calculation using the FRDM model with $Y_{e,0} = 0.25$. 

We found that the DZ31 model predicts a composition whose abundance-averaged properties and emission spectra are very similar to those predicted by the FRDM model. We therefore expect that the different yields found for the DZ31 model will not significantly change $f(t)$ for individual particles. In contrast, the high-$Y_{\rm e,0}$ composition has average compositional properties and emission spectra that depart from the reference case (FRDM, $Y_{\rm e,0} = 0.04$), so we calculate for this composition $f(t)$ of all individual decay products for our fiducial ejecta 
($\mej = 5 \times 10^{-3} \Msun$, $\vej = 0.2c$). 
The thermalization
timescales, plotted in Figure \ref{fig:t50} as
open triangles, are similar to those for the standard
low-$Y_{\rm e,0}$ composition. 
For both the DZ31 and high-$Y_{\rm e,0}$ cases then, impacts on \ftot\ result from differences in the relative importance of each heating channel, not  differences in how efficiently individual decay products
thermalize.

\begin{figure}
\includegraphics[width = 3.5 in]{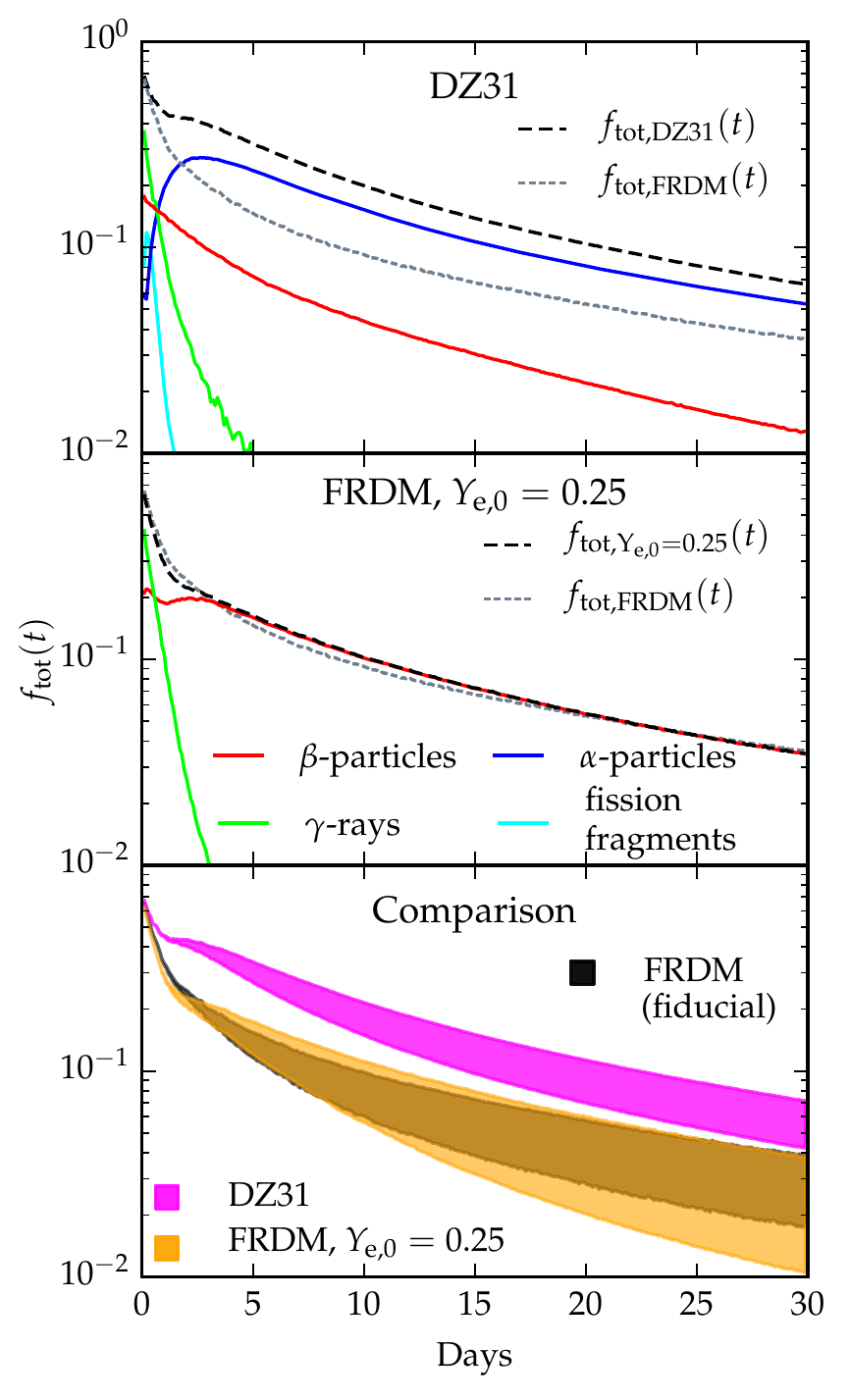}
\caption{The effect of nuclear physics inputs on total thermalization
  efficiency. \textit{Panels 1 and 2:} $f(t)$ (fiducial \mej, \vej;
  random fields) convolved with fractional energy generation rates for the DZ31
  nuclear mass model (top panel) and a high-$Y_{\rm e,0}$ FRDM
  trajectory (middle panel). Solid lines show the fraction of emitted
  energy thermalized by each particle as a function of time, and
  $f_{\rm tot}(t)$ is plotted in black dashed lines. \textit{Panel 3:}
  The range of $f_{\rm tot}(t)$ expected for each of the cases shown
  in Panels 1 and 2. We plot $f_{\rm tot}(t)$ for the low-$Y_{\rm e,0}$ FRDM composition in black for comparison. The widths of the curves are due to
  the range of possible magnetic field configurations.}
\label{fig:ft_nucDep}
\end{figure}

Figure \ref{fig:ft_nucDep} compares \ftot\ for the three cases studied.
In the top panel, we show $f_{\rm tot}(t)$ and the
contributions from each decay product, determined using energy generation rates from
the DZ31 nuclear mass model abundances, for which \al-decay dominates
the energy production at late times.  The middle panel shows an
analogous calculation for the FRDM model with $Y_{\rm e,0} = 0.25$,
which has negligible late-time \al-decay.
In the bottom panel, we compare $f_{\rm tot}(t)$ for these models with
the fiducial FRDM model. The greater role
of \al-decay in the DZ31 model increases \ftot\ by a factor of
$\gtrsim 1.5$, mainly due to the fact that less energy is lost in
neutrinos and \g-rays, which thermalize very inefficiently. In the fiducial composition \al-decay and fission produce only a
small fraction of the energy, so the effect
of increasing $Y_{\rm e,0}$ is modest. A stronger effect might be seen for DZ31, which produces more translead nuclei when $Y_{\rm e,0}$ is low, and is therefore more likely to experience dramatic decreases in translead production when the initial electron fraction rises.

\begin{table}
\centering
\caption{Analytic fit parameters for $f_{\rm tot}(t)$}
\label{tab:coeffs}
\begin{tabular}{cc|ccc}
\multicolumn{2}{c|}{Model} & \multicolumn{3}{c}{Coefficients} \\
$M/M_{\odot}$ & $v_{\rm ej}/c$ & $a$ & $b$ & $d$ \\
\hline
\hline
$1 \times 10^{-3}$ & 0.1 & 2.01 & 0.28 & 1.12\\
$1 \times 10^{-3}$ & 0.2 & 4.52 & 0.62 & 1.39\\
$1 \times 10^{-3}$ & 0.3 & 8.16 & 1.19 & 1.52\\
$5 \times 10^{-3}$ & 0.1 & 0.81 & 0.19 & 0.86\\
$5 \times 10^{-3}$ & 0.2 & 1.90 & 0.28 & 1.21\\
$5 \times 10^{-3}$ & 0.3 & 3.20 & 0.45 & 1.39\\
$1 \times 10^{-2}$ & 0.1 & 0.56 & 0.17 & 0.74\\
$1 \times 10^{-2}$ & 0.2 & 1.31 & 0.21 & 1.13\\
$1 \times 10^{-2}$ & 0.3 & 2.19 & 0.31 & 1.32\\
$5 \times 10^{-2}$ & 0.1 & 0.27 & 0.10 & 0.60\\
$5 \times 10^{-2}$ & 0.2 & 0.55 & 0.13 & 0.90\\
$5 \times 10^{-2}$ & 0.3 & 0.95 & 0.15 & 1.13\\
\end{tabular}
\end{table}

\section{Effect on Kilonova Light curves}
\label{sec:LCs}
To determine the effect of thermalization on kilonova
observables, we incorporated our results for \ftot\ into the
time-dependent Monte Carlo radiation transport code \texttt{Sedona}
\citep{Kasen_MC}, and carried out light curve calculations. The calculations here resemble those of \citet{Barnes_2013}, but include thermalization effects.

\subsection{Analytic fit to thermalization efficiency}

For easy inclusion of thermalization effects in light
curve simulations, we propose a simple analytic formula for $f_{\rm
  tot}(t)$ which provides a good fit to our detailed 
numerical calculations,
\begin{equation}
f_{\rm tot}(t) =  0.36\left[ \exp\left( -at \right) +\\
\frac{\ln \left( 1 + 2bt^d \right) }{2b t^{d}} \right],
\label{eq:fTot}
\end{equation}
where $a$, $b$, and $d$ are fitting constants. The parameterized form
of Eq. \ref{eq:fTot} is motivated by our approximate analytic
solutions for $f(t)$ (Eq.s \ref{eq:ftAn} and \ref{eq:ftg_an}), with
slight modifications to improve the quality of the fit and account for
energy lost to neutrinos. Table~\ref{tab:coeffs} gives the best-fit
parameters for all the ejecta models considered. These fits assume the
FRDM nuclear mass model and random magnetic fields.  

We found that, for the FRDM mass model, compositions from high-$Y_{\rm e,0}$ ejecta have thermalization profiles similar to compositions from initially neutron-rich ejecta. 
This suggests that our thermalization models may be appropriate for material ejected dynamically and from disk winds, regardless of the initial electron fraction. 
However, we note that the insensitivity of \ftot\ to $Y_{\rm e,0}$ may not be as robust for other nuclear mass models. 
The effect of $Y_{\rm e,0}$ may be particularly pronounced for the DZ31 model, which produces large amounts of translead nuclei, and therefore predicts significant \al-decay. Changes in $Y_{\rm e,0}$ could inhibit the production of these nuclei, decrease the role of \al-decay, and thus alter thermalization efficiency.
The effect on \ftot\ would be much stronger than for the FRDM model, which does not produce many translead nuclei even for favorable $Y_{\rm e,0}.$

\subsection{Bolometric light curves}
 
\begin{figure}
\includegraphics[width=3.5 in]{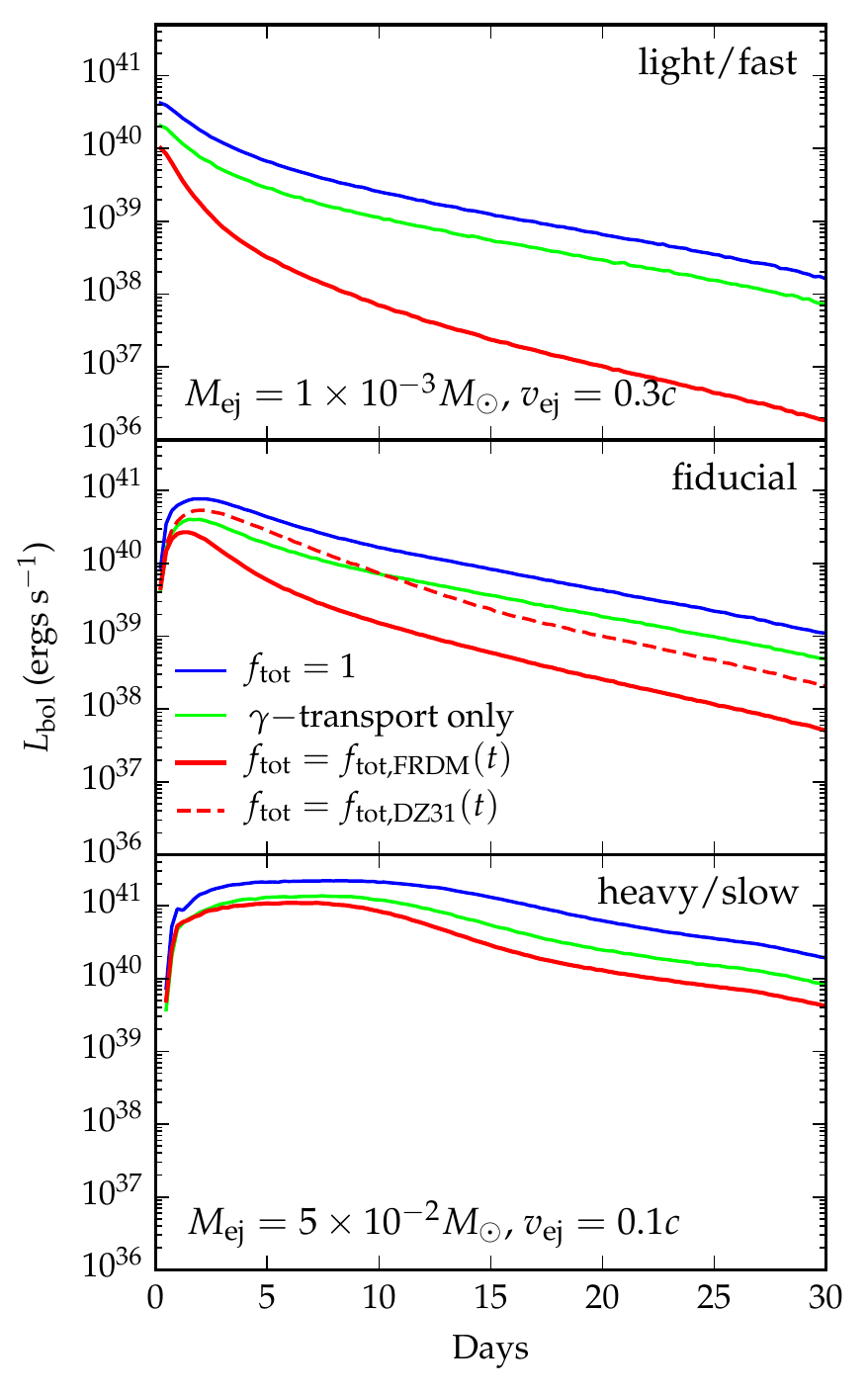}
\caption{Synthetic bolometric light curves for our fiducial ejecta
  model, calculated with \texttt{Sedona} for three different
  treatments of thermalization: full thermalization (blue curve);
  \texttt{Sedona}'s original thermalization scheme, which deposits
  charged particle energy but explicitly tracks the deposition of
  \g-ray energy (lime curve); and the time-dependent $f_{\rm tot}(t)$
  from our numerical simulations (red curve). Accounting for
  time-dependent thermalization efficiencies has a significant impact
  on kilonova luminosity, particularly for models with lower masses
  and higher luminosities. For our fiducial model, the predicted
  luminosity is lower by a factor of $\lesssim 2$ at peak, and by 10
  days is lower by an factor of 5.}
\label{fig:LCeffs}
\end{figure}

The net thermalization efficiency, $f_{\rm tot}(t)$, has a significant
impact on kilonova luminosity.  Figure \ref{fig:LCeffs} compares
bolometric light curves calculated using our derived $f_{\rm tot}(t)$
to those assuming 100\% thermalization. We also show results for a
treatment which propagates \g-rays, but assumes charged particle
energy thermalizes instantly.  This was the method used to estimate
$f_{\rm tot}(t)$ in earlier \texttt{Sedona} kilonova simulations,
including \citet{Barnes_2013}.\footnote{\texttt{Sedona}'s original
  treatment of thermalization assumed that \bt-decay generated 90\% of
  the \rp\ decay energy, with fission accounting for the other 10\%. Of the
  \bt-decay energy, 25\% was taken to be lost to neutrinos, and the
  remaining 75\% was split evenly between \bt-particles and
  \g-rays. The energy from \bt-particles and fission fragments was
  thermalized promptly, while the energy from \g-rays was converted
  into 1 MeV photons, which were propagated through the ejecta in a
  Monte Carlo transport scheme.} (A similar simplification was invoked in the discussion of net heating by \citet{Hotokezaka_2015_heating}.) For all radiation transport
simulations, we have used the simplified composition and the boosted,
synthetic \rp\ opacities of \citet{Kasen_2013_AS}. We consider here
only models with low $Y_{\rm e,0}$, which robustly produce \rp\
elements including Lanthanides and Actinides, making our choice of
opacity appropriate. Models with higher initial electron fractions may
fail to produce these heavy elements. The opacities for such models
would be much lower, and the associated light curves would be shorter,
brighter, and bluer \citep[e.g.][]{Metzger_2010, Barnes_2013, Kasen_Fernandez_Metzger_dwkilonova}.

Figure~\ref{fig:LCeffs} shows that our more accurate treatment of
thermalization impacts predicted photometry for all ejecta models
considered. Relative to earlier calculations with less sophisticated
thermalization schemes, we find kilonova light curves peak
slightly earlier, have lower luminosities at peak, and have much
dimmer late-time luminosities.

The effects of thermalization are most pronounced for less massive and
higher velocity ejecta models, which are dimmer and fade more quickly
than their slower, more massive counterparts.   
  Nuclear physics plays a role by determining the yields of translead
  nuclei, and therefore the amount of energy produced by
  \al-decay. Models for which \al-decay contributes prominently to the
  energy generation yield somewhat higher $f_{\rm tot}(t)$. The middle
  panel of Figure \ref{fig:LCeffs}, which shows light curves
  corresponding to both FRDM and DZ31 (high \al-heating) mass models,
  illustrates this point. The increased luminosity found
    for the DZ31 composition is a result both of higher \ftot\ due to
    more \al-decay, and to the greater absolute amount of energy produced by
    \rp\ decay for the DZ31 model relative to the FRDM model (see \S
    \ref{subsec:Ej_Radio}.)

\subsection{Implications for the kilonova accompanying GRB 130603B}

An excess near infrared (NIR) flux discovered in the afterglow of the
short gamma ray burst GRB 130603B has been widely interpreted as a
kilonova \citep{Tanvir_2013_kNa,
  Berger_2013_kNa}. \citet{Tanvir_2013_kNa} determined that the source
of the flux had an absolute AB magnitude in the \emph{J}-band of
-15.35 at $t \sim 7$ days.  
Having incorporated $f_{\rm tot}(t)$ into kilonova light curve models,
we can more confidently constrain the mass ejected in the kilonova
associated with GRB 130603B. 

\begin{figure}
\includegraphics[width=3.5 in]{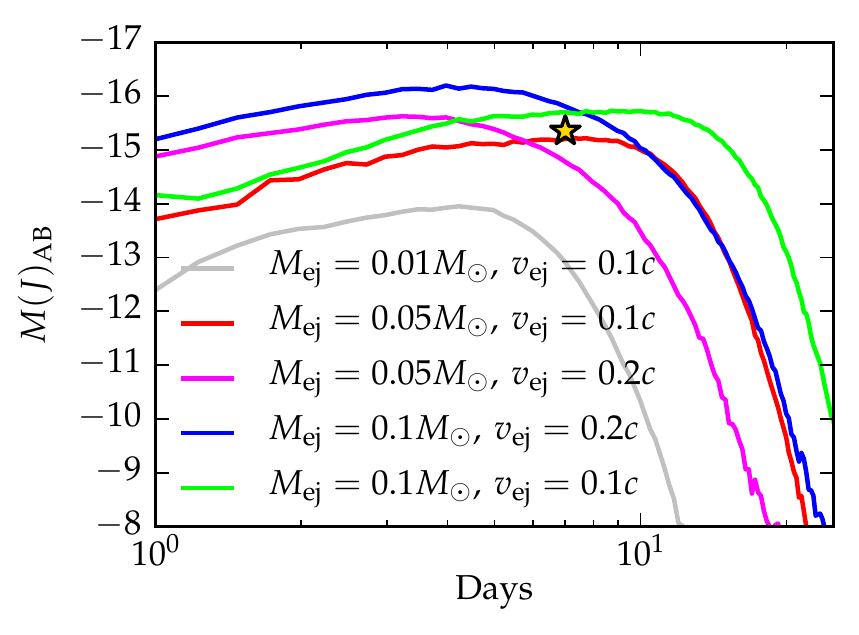}
\caption{Absolute (AB) \emph{J}-band light curves for several ejecta models. The excess IR flux (gold star) suggests an ejected mass between $5 \times 10^{-2}$ and $10^{-1} \Msun$. }
\label{fig:ir_bbs}
\end{figure}

In Figure \ref{fig:ir_bbs}, we compare the detected flux to
\emph{J}-band light curves for various ejecta models, and find the
observed flux is consistent with
$5 \times 10^{-2} M_{\odot} \lesssim M_{\rm ej} \lesssim 10^{-1}
\Msun$.
This mass is higher than what is typically predicted for the dynamical
ejecta from a binary neutron star merger, suggesting that if the
kilonova interpretation is correct, the progenitor of GRB~130603B was
perhaps a neutron star-black hole merger, or that the mass ejected was
significantly enhanced by post-merger disk winds.

Our mass estimate here is an improvement over earlier work which neglected detailed thermalization, and gives substantially different results. For example, \citet{Piran_2014_130603BkNa} suggested $\mej \sim 0.02 \Msun$, less than half our new value.
However,
we have not accounted for viewing angle effects. 
If the ejected material is mainly confined to the equatorial
plane, the emission will be brighter when the system is viewed face-on \citep{Roberts_2011}, which
would reduce the inferred mass somewhat. If the ejecta is oblate, thermalization will also be more efficient, which could have a small impact on mass estimates. Radiation transport simulations in three dimensions with time-dependent thermalization models will further constrain $M_{\rm ej}$.

\subsection{Late-time light curve}
\label{subsec:lateLC}
Late time kilonova light curves may probe the history of  \rp\
nucleosynthesis in CO mergers. At $\sim 2$ days after merger, fission
ceases to be important, and \al- and \bt-decay dominate the kilonova's
energy supply. Energy from \al-decay is transferred entirely to fast
\al-particles, which thermalize fairly efficiently out to late
times. Beta particles thermalize with similar efficiency, but carry
only a fraction ($\sim 25\%$) of the total \bt-decay energy, with the
rest lost to neutrinos and \g-rays. A kilonova's late-time luminosity
will therefore depend on the relative importance of \al- versus
\bt-decay. Because only nuclei with $200 \lesssim A \lesssim 250$
undergo \al-decay, the late time kilonova luminosity may diagnose the
presence of heavy elements in the ejecta, and therefore constrain the
neutron-rich conditions required for heavy element formation. 

\begin{figure}
\includegraphics[width=3.5 in]{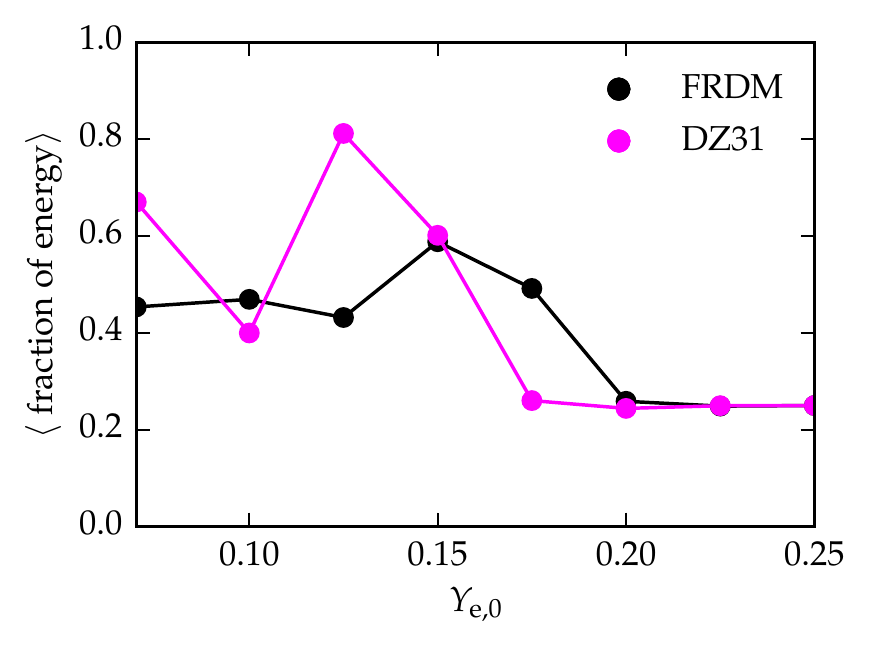}
\caption{The time-averaged fraction of the \rp\ decay energy emitted as \bt- and \al-particles and fission, as a function of initial $Y_{\rm e}$. The data was taken from \rp\ calculations on our representative SPH trajectory, with $Y_{\rm e,0}$ artificially altered.  The lower the initial electron fraction is, the more energy from \al-decays is available to drive a late time light curve.
}
\label{fig:ltLCedot}
\end{figure}

We gauge the relative strength of late-time kilonova light curves for
different $Y_{\rm e,0}$ by estimating the percent of energy from the decay of \rp\ elements
emitted as fission fragments, \al-, and \bt-particles, time-averaged over $t=10-100$
days. (Note that while all energy from \al-decay emerges as \al-particles, \bt-\emph{particles} receive only 25-30\% of the energy from \bt-\emph{decay}.) The results for our representative SPH trajectory, for a range
of $Y_{e,0}$ and two nuclear mass models, are shown in Figure
\ref{fig:ltLCedot}.  The curves suggest that systems with
$Y_{\rm e,0} \lesssim 0.17$ have more robust late-time heating, and are likely to exhibit late time light curves that are more luminous by a factor of up to $\sim 2$.

If fission is more significant at late times than our calculation predicts (e.g. \citet{Hotokezaka_2015_heating} find that fission supplies $\gtrsim 10\%$ of the total energy out to late times) the dependence of the late-time light curve on $Y_{\rm e,0}$ could be much stronger. Fission fragments thermalize extremely efficiently well past maximum light. Since very neutron-rich conditions are needed to build up the heavy nuclei ($A \gtrsim 250$) that undergo fission, in a strongly fissioning ejecta, late-time luminosity could depend sensitively on $Y_{\rm e,0}$.

A key observable for probing the nuclear physics of 
CO mergers may therefore be the ratio of the kilonova luminosity measured at peak brightness to that observed
on the light curve tail. The former is powered by $\beta$-decay, while the latter is driven by $\alpha$-decay and fission, which thermalize more efficiently.  This ratio could therefore constrain the composition of the ejecta, and so the conditions of nucleosynthesis.  More detailed studies are needed to clarify this relationship.

\section{Conclusion}

We have shown that the radioactive energy from the decay of \rp\ material does not completely thermalize  in the ejecta from CO mergers. For the first time, we have explicitly simulated the thermalization of all suprathermal \rp\ decay products in the heavy-element-rich kilonova ejecta. From these simulations, we derived time-dependent expressions for the net kilonova thermalization efficiency, $f_{\rm tot}(t)$, for models spanning a range of expected ejecta masses and velocities. For most parameters studied, \ftot\ drops below 0.5 within five days after the merger.  At 15 days after merger, \ftot\ may be as low as $0.01-0.1$.  Thermalization therefore has a significant impact on the peak luminosity of kilonovae and the late time light curve decline rate.

We have also explored the dependence of $f_{\rm tot}(t)$ on electron fraction and nuclear mass model, and outlined how variations in these parameters may systematically
affect thermalization. In general, systems that favor the production of translead nuclei have higher thermalization efficiencies at late times.  This is because a greater fraction of radioactive energy is emitted through $\alpha$-decay and fission channels, which thermalize more efficiently than energy from $\beta$-decays, since less energy is lost to \g-rays and neutrinos.

We have presented updated  radiation transport simulations that incorporate our new calculations of time-dependent thermalization, and find that thermalization has a significant effect on the predicted photometry of kilonovae. Compared to older models that neglected detailed thermalization, our new light curves peak earlier and at lower luminosities, and are much dimmer (by a factor of $\sim 10$) at late times ($\gtrsim 15$~days after peak). Our new models of kilonova with Lanthanide-rich \rp\ ejecta keep their characteristic red color, and are much more luminous in infrared (\emph{I}-, \emph{J}-, and \emph{K}-) bands than in the optical.

Our results have consequences for detecting kilonovae, whether
blindly, or as counterparts to gravitational wave events. While our
new models retain the red color believed to be a defining kilonova
signature, the rapid decline of \ftot\ poses a challenge to detection,
and underscores the necessity of timely follow-up of GW
triggers. Since \ftot\ is lower, and declines more quickly, in less
massive systems, this is especially important for kilonovae generated
by merging neutron stars, which are expected to be less massive than
BHNS kilonovae by a factor of $\sim 10$.

The recent detection of the gravitational wave event GW150914
\citep{LIGO_150914} spurred a slew of EM follow-up activities
\citep{EM_followup_150914,DeCAM_Followup_GW150914}. While the GW
signal turned out to be the result of a binary black hole merger,
which is not expected to have an EM counterpart, the follow-up
campaign offers a sense of the prospects of detecting a kilonova
counterpart to future GW events. From a kilonova standpoint, some of
the most promising observational efforts were carried out by the Dark
Energy Camera (DECam), which had limiting (AB) magnitudes $i < 22.5$
and $z < 21.5$; the VLT Survey Telescope (VLS), which reached
$r < 22.4$, and VISTA, with $J < 20.7$.

Had the GW trigger been due to a typical NSBH merger located 100 Mpc
distant, the associated kilonova ($\mej = 5 \times 10^{-2} \Msun$;
$\vej = 0.2 c$) could have been observed by DECam in \emph{i} and
\emph{z} for $t \lesssim 7$ days; by VST in \emph{r} for
$t \lesssim 5$ days; and by VISTA in \emph{J} for $t \lesssim 8$
days. The situation is less promising for a NS$^2$ merger ejecting
less mass ($\mej = 5 \times 10^{-3} \Msun$; $\vej = 0.2 c$),
which will be intrinsically dimmer and suffer from less efficient
thermalization. To be visible to DECam in \emph{i} (\emph{z}) at peak,
such a system would need to be closer than $\sim 63$ ($\sim 57$) Mpc,
while VISTA (VST) could only detect it at distances less than
$\sim 52$ ($\sim 10$) Mpc.

This analysis highlights the importance of seeking optical
counterparts at early times, before they fade below detection
thresholds. It also suggests that observing strategies should focus on
depth, rather than area, to improve the chances of detecting signals
that are likely to be faint. Lastly, these findings emphasize the
criticality of developing facilities with greater IR
sensitivity. \emph{Euclid} \citep{Amendola_2012_Euclid} and
\emph{WFIRST} \citep{Green_2012_WFIRST}, each which will have an
\emph{H}-band depth of $\sim 25$, could detect a typical NS$^2$
kilonova, located at 100 Mpc, out to $t \sim 15$ days.

Our calculation of time-dependent thermalization efficiencies for
kilonovae constrains a key uncertainty in models of \rp\
transients. Additional work can further improve these models. We have
focused on thermalization in the dynamical ejecta from compact object
mergers, but the ideas developed here could---and should---be applied to
study disk wind outflows, which may produce ejecta poor in
  Lanthanides and Actinides and potentially contribute a
  blue/optical component to kilonova light curves. Calculating
heating efficiencies for the multiple components believed to make up a
kilonova, and incorporating realistic models of \ftot\ in
three-dimensional radiation transport simulations of multi-component
light curves would yield the best predictions to date of kilonovae's
EM signatures.

\acknowledgments

This work is supported in part by a Department of Energy Office of Nuclear
Physics Early Career Award, and by the Director, Office of Energy
Research, Office of High Energy and Nuclear Physics, Divisions of
Nuclear Physics, of the U.S. Department of Energy under Contract No.
DE-AC02-05CH11231, and from NSF grant AST-1206097
Support for M-RW and GM-P is provided in part by the Helmholtz Association through the
Nuclear Astrophysics Virtual Institute (VH-VI-417) and the
BMBF-Verbundforschungsprojekt number 05P15RDFN1.

\appendix
\section{Transport methods}
\label{sec:methods}
This appendix describes our numerical method for calculating detailed 
thermalization efficiencies using a 3-dimensional (3D) particle transport scheme.

\subsection{Grid}
We carry out our calculation on a 3D Cartesian grid of $100^3$ 
zones.  The ejecta is assumed to have a broken power-law density profile and a two-zone composition, as described in \S \ref{sec:ejMod}. The grid is initialized at $t=0.1$ days, and advanced in time every 0.1 days.

In each time step, the grid emits $N_{\rm pack}$ particle packets with co-moving frame (cmf) kinetic energy $E_{\rm pack}$ equal to $E_{\Delta t}(t)/N_{\rm pack}$, where $E_{\Delta t} = \mej\int_{\Delta t} \dot{\epsilon}(t) \dd{t}$ is the total radioactive energy emitted by the ejecta over the course of the time step. 
The probability of emission in any zone is proportional to zone mass.
Packets are emitted at random positions within the zone, with random initial directions, at a time selected from a flat distribution $[t, t + \Delta t)$.
Each packet represents $N_{\rm part}$ particles of energy $E_{\rm part}$, where the particle's initial cmf energy is sampled from the time-dependent spectra of Figure \ref{fig:netSpec}.
Particles are transported through the grid until
their cmf energy falls below a threshold energy $\sim 1000$ times less than the typical energy at emission. At this point, particles deposit their residual kinetic energy and are removed from the grid.

\subsection{Gamma rays}
Gamma ray packets are propagated through the grid in small steps in typical Monte Carlo fashion
(e.g. \citealt{Lucy_2005_MC, Kasen_MC}.)
A propagating photon packet can reach the end of the current grid time step, leave the zone it is in, or interact with an atom. The outcome that occurs on the shortest timescale is selected.
The timescales for the first two processes are straightforward, and the timescale for interaction is given by
\begin{equation}
\Delta t_{\rm atom} = -\frac{1}{c \rho \kappa_{\rm tot}}\ln(1 - z),
\end{equation}
where $\kappa_{\rm tot} = \kappa_{\rm C} + \kappa_{\rm PI}$ is the sum of of Compton and photoionization opacities, $\rho$ is the mass density, and $z$ is a random number sampled from $[0,1)$.
The relative weights of $\kappa_{\rm C}$ and $\kappa_{\rm PI}$ determine whether the \g-ray packet Compton scatters or photoionizes.

During a Compton scatter, a photon  undergoes some angular displacement, with the differential cross section and change in energy given by the Klein-Nishina formula. Energy lost by the photon packet is transferred to a non-thermal electron, which also acquires a momentum opposite to that of the post-scatter photon. The electron is then transported through the grid, as described in the next section.

In a photoionization, some of the photon's energy thermalizes promptly, but the rest is transferred to a non-thermal photoelectron. In general, photoelectrons are ejected from the inner shells of the ionized atoms, and acquire energies equal to the \g-ray energy less the photoelectron's binding energy, $E_{\rm B}$. Outer shell atomic electrons cascade down to fill the vacancy created by the ionization, releasing photons whose energies sum to $ \sum E = E_{\rm B} - \Delta E_{\rm atom}$, where $\Delta E_{\rm atom}$ is the difference in the atom's total energy before and after the ionization and cascade. The ejected photoelectron will travel through the grid until it thermalizes and recombines with an ambient atom. Since the thermalized electron will attach to the atom's valance shell, the energy released upon recombination will be low, of order $\Delta E_{\rm atom}$.

We assume that all photoelectrons originate from inner shells of heavy elements, so $E_{\rm B} \gg \Delta E_{\rm atom}$. The energy released in the post-ionization cascade is then $E_{\rm B} - E_{\rm atom} \approx E_{\rm B}$. This energy thermalizes immediately. The photoelectron receives the remainder of the \g-ray energy, $E_\g - E_{\rm B}$, and a momentum parallel to the initial \g-ray momentum. It is propagated until it dissipates its kinetic energy.

Since the cross section for photoionization peaks at $E_{\g} \sim E_{\rm B}$, we estimate representative binding energies for the \rp\ composition by locating local maxima in the aggregate photoionization opacity curve. For each photoionization, we select the minimum binding energy such that $E_{\gamma}^{\rm cmf} > E_B$. If $E_{\gamma}^{\rm cmf} \gtrsim 0.1$ MeV, and there are no identifiable peaks that satisfy this criterion, we set $E_{\rm B} = 0.9 E_{\gamma}^{\rm cmf}$. 

\subsection{Massive particles}
\label{subsec:partTrans}
Massive particles lose energy continuously, so are not suited to a discrete Monte Carlo treatment. Instead, a massive particle packet is transported through the grid until it reaches the border of its zone, the grid time step ends, or (in the case of disordered fields) the magnetic field changes direction. The change in the particle's cmf energy is calculated from the particle's initial cmf energy, and the zone conditions and elapsed time, both measured in the cmf. The particle energy is then adjusted.

For \al-particles and fission fragments, all energy lost is assumed to be thermalized and promptly deposited on the grid.
In the case of \bt-particles, energy loss via Bremsstrahlung can be significant, and a non-negligible fraction of the \bt-particle energy can be converted to photons of fairly high energies, which will not immediately thermalize. To account for this possibility, we differentiate \bt-particle energy lost to plasma and ionization/excitation---which we assume thermalizes instantly---from energy lost to Bremsstrahlung, which we assume does not. 
At the end of each time step, thermalized energy $\Delta E_{\rm th}$ is deposited in the grid, and $E_{\rm part}$ and $E_{\rm pack}$ are adjusted accordingly. The probability that the \bt-packet is converted to a \g-packet is given by the ratio of the energy lost to Bremsstrahlung during the preceding time step to the particle energy after thermalization: $P_{\bt \rightarrow \g} = \Delta E_{\rm Br}/(E_{\rm \beta,i}^{\rm cmf} - \Delta E_{\rm th})$, and random number sampling determines the outcome.
If the packet remains a \bt-packet, $E_{\rm part}$ is decremented by $\Delta E_{\rm Br}$, while $E_{\rm pack}$ remains the same (i.e., $N_{\rm pack}$ is updated.)
If conversion to a \g-packet is selected, a new direction of propagation is set randomly in the co-moving frame. The cmf energy of the new \g-ray is selected from a flat probability distribution in the range $(0, E_{\rm \beta,i}^{\rm cmf} - \Delta E_{\rm th}]$, which approximates the fairly flat Bremsstrahlung spectrum of relativistic \bt-particles \citep{Jackson_EM}. The value of $N_{\rm pack}$ is then adjusted to preserve total post-thermalization packet energy.

\subsubsection{Influence of magnetic fields}
Charged particles have Larmor radii much smaller than the coherence length of the magnetic field, so we track their motion along field lines without resolving oscillations about the guiding center. The motion of a particle with mass $m$ and kinetic energy $E$ can then be described by the average velocity
\begin{equation}
\langle \mathbf{u} \rangle = v(E)\mu\mathbf{\hat{B}},
\end{equation}
where $v$ is the total velocity of a particle with kinetic energy $E$, $\hat{\mathbf{B}}$ is the unit vector directed along $\mathbf{B}$, and $\mu$ is the cosine of the angle (the ``pitch angle") between the particle's total velocity and the velocity of its guiding center, which is aligned with $\mathbf{B}$.
This $\langle \mathbf{u} \rangle$ is the particle velocity we use to boost between the co-moving and center-of-explosion frames, and to update particle position at the end of a time step.

We explore three classes of magnetic fields: radial ($\mathbf{B} \propto \mathbf{\hat{r}}$), toroidal ($\mathbf{B} \propto \hat{\boldsymbol{\phi}}$), and random. For radial and toroidal fields, the magnetic field direction at any location in the ejecta can be determined trivially, and updating the direction of $\langle \mathbf{u} \rangle$ at the end of each time step is straightforward. 

For tangled fields, we randomly choose a unique $\hat{\mathbf{B}}$ for each particle upon emission. We assume the field changes on a length scale $\lambda R_{ej}(t)$, where $\lambda$ is a model parameter, and at the beginning of each time step we calculate the timescale for the particle to experience a significant change in $\hat{\mathbf{B}}$:
\begin{equation}
\Delta t_{\rm B} = \frac{\lambda R_{\rm ej}(t)}{|\langle \mathbf{u} \rangle |}\ln(1-z),
\end{equation}
where $z$ is a random number between 0 and 1. If $\Delta t_{\rm B}$ is less than the time for a particle to leave a zone or end the grid time step, the magnetic ``scattering'' action is selected, the particle position and energy are updated, and a new field direction is chosen at random. The re-orientation preserves energy and the magnitude of the momentum in the cmf. This is reasonable, because these discrete scatters are standing in for smoother and more gradual field gradients. However, it also leads, on average, to energy losses in the center of explosion frame, which decrease thermalization efficiency. 
This can be thought off as the particles transferring their energy to the bulk kinetic energy of the ejecta, rather than the thermal background. It is a small effect.

The propagation of a particle along field lines depends on total velocity and $\mu.$ Particles are emitted on the grid with a random direction defined by the unit vector $\hat{\mathbf{D}}$, 
so $\mu$ is $\hat{\mathbf{D}}\cdot \hat{\mathbf{B}}$. 
In the case of radial fields, $\mu$ evolves to preserve the magnetic adiabatic invariant $(1-\mu^2)/B$. This evolution encourages outward motion.
Inward-moving particles encounter ever-stronger fields, and decrease their inward velocity in response, eventually ``mirroring'' off the field and beginning to travel outward. Particles streaming outward move into weaker magnetic fields, so their pitch angle increases and they stream out even faster. 

We include magnetic beaming and mirroring only for radial fields, where changes to the pitch angle facilitate particle escape and can be meaningful for thermalization. Particles in toroidal fields are confined to one region of velocity space; changes in their pitch angle cannot promote outward motion and so cannot affect thermalization. While particles in random fields could experience beaming and mirroring, they may also undergo pitch angle scattering---interactions with small fluctuations in the field that can change pitch angle, and would act to isotropize $\mu$, counter to beaming and mirroring. We assume the two effects balance out, and hold pitch angle constant for particles in random and toroidal fields.

\subsubsection{Fission fragment transport}
\label{subsec:fragTrans}
As discussed in \S \ref{subsec:Bej}, at late times, fission fragments may have Larmor radii comparable to the magnetic field coherence length. This could affect the transport, especially for disordered fields, where field lines in close proximity may have very different orientations. In such a system, the guiding center approximation breaks down, and the motion of the particle must be resolved.
We estimate the importance of this effect by modeling fission fragment transport in a tangled field as a random walk of the fragment itself (as opposed to its guiding center). In this simple scheme, fission fragments travel in straight lines and re-orient randomly on length scales of $r_{\rm L} = r^{\rm ff}_{\rm L,max}= 1.0\times 10^{-2} v_2 t_{\rm d} R_{\rm ej}$. This path stands in for the more complicated looping trajectory we would see when $r_{\rm L} \sim \lambda R_{\rm ej}$. We carried out a simulation of fission fragment transport in this limit for the fiducial ejecta model, and found only negligible variation in $f_{\rm ff}(t)$ relative to the flux-tube approximation. The difference was was apparent only at late times, when fission contributes little to the energy generation. We therefore conclude that detailed models of fission fragment transport are unnecessary.

\hfill
\bibliography{refs}

\end{document}